\documentclass{article}
\usepackage{graphicx} 
\usepackage{booktabs}
\usepackage{hyperref}
\usepackage{rotating}
\usepackage{xltabular}
\usepackage{authblk}
\usepackage{tikz}
\usetikzlibrary{positioning}
\usetikzlibrary{arrows.meta}
\usetikzlibrary{shapes.multipart, positioning, fit, backgrounds}
\usepackage{float}
\usepackage{tabularx}
\usepackage{ragged2e} 
\usepackage{pdflscape}   
\usepackage{booktabs}    
\usepackage{array}       
\usepackage{graphicx}    
\usepackage{graphicx}
\usepackage{booktabs}
\usepackage{float}
\usepackage{afterpage}
\usepackage{caption}
\usepackage[table]{xcolor}
\usepackage{multirow}
\usepackage{booktabs}

\title{Taxonomy-Driven Analysis of Open-Source AI Risk Mitigation Tools}

\author{
Afreen Alam$^{3}$,
Evgenija Popchanovska$^{2}$,
Ana Gjorgjevikj$^{2}$,
Maryan Rizinski$^{1}$,
Lubomir T. Chitkushev$^{1}$,
Irena Vodenska$^{3}$,
Dimitar Trajanov$^{2,1}$\\
\normalsize $^{1}$Department of Computer Science,\\
\normalsize Metropolitan College, Boston University, Boston, MA 02215, USA\\
\normalsize $^{2}$Faculty of Computer Science and Engineering,\\
\normalsize Ss. Cyril and Methodius University, Skopje 1000, North Macedonia\\
\normalsize $^{3}$Department of Administrative Sciences,\\
\normalsize Metropolitan College, Boston University, Boston, MA 02215, USA\\

\normalsize\textbf{Emails:}\\
\normalsize
Afreen Alam: \texttt{afreen99@bu.edu}\\
Evgenija Popchanovska: \texttt{evgenija.popchanovska@students.finki.ukim.mk}\\
Ana Gjorgjevikj: \texttt{ana.gjorgjevikj@students.finki.ukim.mk}\\
Maryan Rizinski: \texttt{rizinski@bu.edu}\\
Lubomir T. Chitkushev: \texttt{ltc@bu.edu}\\
Irena Vodenska: \texttt{vodenska@bu.edu}\\
Dimitar Trajanov: \texttt{dimitar.trajanov@finki.ukim.mk}
}

\date{}

\begin{document}

\maketitle

\begin{abstract}
    The rapid adoption of large language models (LLMs) in enterprise settings has introduced a range of operational, security, and governance risks. As organizations move generative AI applications from pilot deployments to production systems, manual approaches to harm identification and mitigation are becoming increasingly difficult to scale. Although many tools now support tasks such as model evaluation, adversarial testing, runtime guardrails, and observability, the current tooling landscape remains fragmented. Tools are typically built for specific engineering use cases and described in technical terms that do not align cleanly with governance frameworks or risk taxonomies, creating uncertainty about which tools address which risks, where capabilities overlap, and where critical gaps remain. This paper proposes a structured protocol to automate AI risk mitigation through a taxonomy‑driven analysis of open‑source LLM evaluation and security tools. We systematically map the technical capabilities of 21 prominent open‑source tools to the 32 sub‑categories of the extended MIT AI Risk Mitigation and Response Taxonomy. An LLM‑assisted retrieval‑augmented generation pipeline is used to analyze the relevant source code and documentation for each tool and extract relevant capabilities mapping to each taxonomy category. The reliability assessment of these results yielded a moderate agreement of 0.509, measured using Fleiss’ Kappa, between three independent reviewers. The analysis reveals a highly skewed landscape in which open‑source tools clustered around technical and operational controls (e.g. model safety engineering, content safety, data governance, and post‑deployment monitoring), while governance oversight, legal and regulatory remedies, and financial and market controls remain largely unaddressed by code‑level mechanisms. This motivates a layered risk‑mitigation architecture proposal in which tool‑based controls are explicitly complemented by organizational and regulatory processes. The LLM‑assisted mapping protocol achieves reliable performance in this interpretive classification task, with a F1 score of 75.5\% post majority vote resolution between the three reviewers. In general, the study offers a practical mapping between enterprise AI risk categories and open‑source mitigation capabilities, and identifies areas where automated tooling alone is insufficient and human oversight remains necessary. Although the analysis focuses on open‑source tools, the taxonomy‑driven framework is designed to be applicable to proprietary solutions that organizations can incorporate into their workflows.

\end{abstract}

\section{Introduction}
Large language models (LLMs) are increasingly being deployed in enterprise environments to support decision-making, customer interaction, software development, knowledge retrieval, and internal business analytics. However, their adoption introduces risks that differ from those associated with conventional software systems. Because LLMs generate probabilistic results, they can produce inaccurate or unsupported claims, expose sensitive information, respond unpredictably to adversarial inputs, or reproduce biased patterns embedded in training data \cite{anwar2024foundationalchallengesassuringalignment, broadwater2026evaluatingreliabilitygapslarge}. These risks are especially consequential in regulated sectors such as financial services, insurance and legal services, where system failures can lead to compliance violations, financial loss, reputational harm, or adverse effects on individuals \cite{lseg2024llmsfinance, Desai_2024, bis2025fsi}. As enterprise adoption of generative AI accelerates, organizations require standardized, multi-metric evaluation and reliable uncertainty quantification to support technically effective integration and trust \cite{liang2023holisticevaluationlanguagemodels}.

Manual review processes may be appropriate during early experimentation, but are rarely sufficient for production systems that process large volumes of user interactions in real time. Moreover, risk exposure is not limited to model behavior alone. It also arises from system design choices, retrieval pipelines, prompt handling, operational incident monitoring, access controls, and downstream integration with enterprise workflows \cite{anwar2024foundationalchallengesassuringalignment,fsb2025monitoring}. As a result, effective governance requires a combination of technical controls, organizational procedures, and ongoing monitoring.

A growing ecosystem of open-source tools has emerged to address these challenges. These tools provide capabilities such as prompt-injection testing, red teaming, hallucination detection, personally identifiable information filtering, policy enforcement, model evaluation, and observability. However, the ecosystem remains difficult for enterprises to navigate and interpret. Tools are often developed for specific engineering use cases and described using technical terminology that does not align directly with governance frameworks or academic risk taxonomies. Consequently, organizations may struggle to determine which tools address which risks, where tool capabilities overlap and where important gaps remain. Financial authorities and supervisors, in particular, may struggle to monitor AI adoption and its risks due to significant uncertainty from rapid innovation and limited data on AI uptake \cite{bis2025fsi, liang2023holisticevaluationlanguagemodels}. 

This paper aims to address this by systematically mapping open-source LLM evaluation and security tools to a structured AI risk mitigation taxonomy.  Rather than treating tools as isolated developer utilities, we examine how their documented and executable capabilities correspond to recognized risk categories originally compiled by MIT and later extended in \cite{popchanovska2026aifailsworksdatadriven} (see Appendix A for a full listing of taxonomy categories and subcategories). The goal is to provide a clearer basis for enterprise architecture decisions, tool selection, and automated risk mitigation design. All resources pertaining to the implementation of this mapping has been made publicly available and can be accessed at the paper's Github repository link\footnote{https://github.com/afreen99/ShieldAI-A-Taxonomy-Driven-Analysis-of-Open-Source-AI-Risk-Mitigation-Tools/tree/main}. 

The paper proceeds as follows. Section 2 reviews the relevant literature on AI governance frameworks, LLM safety evaluation, open-source security tooling, and the gap between academic taxonomies and engineering practice. Section 3 presents the methodology, including tool selection, repository ingestion, LLM-assisted capability extraction, human validation, and consensus adjudication. Section 4 presents the landscape analysis and key empirical findings along with a proposal of an enterprise risk mitigation architecture that combines evaluation, runtime guardrails, observability, and human oversight. Section 5 outlines opportunities for extending this work, including scenario-based evaluation of layered tool stacks and extension of the mapping methodology to proprietary tools, alongside the study's key limitations. The paper concludes by discussing the limitations of software-based mitigation and identifying directions for future research.

\section{Related Work} 
Recent work has begun to document the methodological and practical limits of current LLM evaluation practices, but it remains largely disconnected from the governance and risk‑management needs of enterprises, particularly in high‑stakes domains such as financial services. Position papers on LLM safety evaluation argue that existing assessments are highly fragmented and often lack robustness, noting that results hinge on small, ad‑hoc datasets, inconsistent implementations, and heterogeneous pipelines that make safety claims difficult to validate and LLM evaluations difficult to properly implement. At the same time, emerging benchmark suites propose more systematic reliability and security testing regimes, but they largely center on model‑level evaluation, datasets, and judge standardization rather than on organizational risk‑management processes \cite{beyer2026llmsafetyevaluationslackrobustness}.

A strand of work focused on industry deployments has responded by characterizing the specific challenges that arise when LLMs are used in regulated or economically sensitive settings. The use of LLM in finance, for example, provides a comprehensive overview of applications such as report generation, market‑trend forecasting, sentiment analysis, and personalized advice, but emphasizes that ensuring precision, reliability, and regulatory compliance remains a “major challenge” due to domain‑specific data constraints and strict oversight requirements. Broader surveys of generative‑AI in finance similarly highlight challenges associated with scarce high‑quality private data, the costs and limitations of fine‑tuning, inference latency, and deployment complexity, and the heightened consequences of hallucinations and inconsistent reasoning, which can directly translate into financial and reputational loss or regulatory breaches. Industry‑oriented analyses of LLM risks in finance highlight biased decision‑making, accuracy and reliability concerns, regulatory non‑compliance, cybersecurity vulnerabilities, and privacy risks as major issues that require careful model‑risk assessment, measures to ensure compliance and responsible use, and a combination of expert systems and manual review mechanisms to address them \cite{zhao2025revolutionizingfinancellmsoverview,lseg2024llmsfinance}.

More broadly, LLM safety and alignment surveys emphasize the structural limitations of current mitigation strategies, especially the overreliance on fine‑tuning as a one‑time intervention rather than part of an ongoing lifecycle. These works note that models often behave unpredictably under distribution shift, adversarial prompting, or long‑horizon interaction, even when they appear safe on curated benchmarks. Subsequent empirical studies examining safety under repeated inference further demonstrate that models can pass single‑shot evaluations while still exhibiting significant reliability gaps when queried at scale or over extended periods. Together, this literature points to a mismatch between one‑off pre‑deployment evaluations and the probabilistic, evolving nature of harms that emerge as systems move from pilot environments to live production use, highlighting how static assessments and fine‑tuning are insufficient to assure safety over time and can introduce forward‑looking bias and other tensions with established model‑risk‑management expectations in financial institutions \cite{anwar2024foundationalchallengesassuringalignment,lseg2024llmsfinance}.

In parallel, several initiatives have sought to bring more structure and standardization to LLM assessment in ways that are explicitly intended to be useful to practitioners. The Holistic Evaluation of Language Models (HELM) framework advocates transparent and reproducible benchmarking across tasks and scenarios, with an emphasis on covering a broad space of applications and documenting evaluation conditions. The NIST AI Risk Management Framework (AI RMF 1.0) and its Generative AI Profile (NIST AI 600-1) extend this trend beyond model-centric evaluation by articulating a lifecycle-oriented approach to AI risk management around four core functions, Govern, Map, Measure and Manage, that structure activities from policy design through measurement and continuous improvement in socio-technical systems \cite{NIST-AI-RMF, NIST-GEN-AI}. Although these frameworks represent important steps toward a more systematic evaluation landscape, HELM focuses on taxonomizing scenarios and metrics and on standardized benchmarking rather than specifying how evaluation outputs should be mapped onto organizational governance structures and risk taxonomies \cite{liang2023holisticevaluationlanguagemodels}.

A small but growing body of work begins to consider deployment‑specific and risk‑aware behavior more explicitly in complex operational settings. Research on risk‑aware LLM agents for specialized domains integrates safeguards directly into agentic workflows and discusses practices for production‑grade deployment, illustrating how technical architectures can be aligned with domain‑specific risk profiles. Production‑oriented benchmark suites such as Swiss‑Bench 003 similarly emphasize reliability and security under realistic usage patterns, including repeated evaluation and adversarial probing, rather than single‑point estimates of performance. In the financial sector specifically, surveys of LLM applications highlight that most deployments to date remain in relatively low‑stakes or support functions, precisely because institutions lack confidence that current evaluation and monitoring practices are sufficient for higher‑impact use cases like credit decisioning, trading, or financial crime detection. These contributions illustrate how technical tools and research can be oriented towards operational risk reduction in finance, ddressing trustworthiness, interpretability, and regulatory challenges, but many mitigation techniques remain emerging, are often targeted to specific domains or evaluation dimensions, struggle to generalize across diverse financial tasks, and underscore the need for further work on safe, effective, and responsible deployment and on orchestrating multiple tools across the model lifecycle \cite{llmsfinance2025survey,zhao2025revolutionizingfinancellmsoverview,lseg2024llmsfinance}.

Finally, work on post‑deployment monitoring and AI governance highlights a complementary set of organizational and regulatory challenges that are especially visible in finance. Analyses of post‑deployment monitoring practices argue that current mechanisms are underdeveloped, citing limited transparency, restricted access for external scrutiny, and weak internal structures for ongoing oversight, and call for standards and infrastructure that enable continuous monitoring and clear lines of accountability. The Treasury guidance emphasizes integration of AI risk in existing risk and compliance frameworks and provides interrelated deliverables to support governance, but it does not prescribe how institutions should select or map specific technical tools, benchmarks or monitoring platforms into concrete risk taxonomies \cite{grantthornton2026treasury}. The FS‑AI RMF similarly notes the need for practical tools and reference materials to help bridge the gap between a growing ecosystem of technical tools and the high‑level governance aspirations that enterprises are expected to operationalize \cite{treasury2026fsairmf}. Recent guidance from the Financial Stability Board on sound practices for responsible AI adoption in financial services likewise recommends continuous monitoring, regular validation, and institution‑wide governance structures as core elements of effective AI risk management \cite{fsb2026soundpractices}. The result is a pronounced gap between the current state of post-deployment monitoring and reporting — including limited transparency, restricted access for external scrutiny, and a lack of standardised technical infrastructure — and the high-level governance aspirations and risk taxonomies that enterprises are expected to operationalize \cite{mokander2022conformity}.

Together, this literature paints a picture of an evaluation and governance landscape that is conceptually rich but operationally fragmented. We see detailed critiques of existing safety evaluations, finance‑specific surveys of applications and challenges, and emerging standards and benchmarks, but relatively little work that systematically maps concrete tool capabilities to enterprise risk categories or to lifecycle‑wide mitigation workflows. In particular, there is limited guidance on how organizations should select and compose open‑source evaluation and security tools to cover specific risk subcategories, how to identify coverage gaps, and how to connect these technical assets to continuous feedback loops that include human oversight and align with sectoral frameworks such as the FS‑AI RMF. This gap motivates the present study, which explicitly links a curated set of open‑source LLM evaluation and security tools to a comprehensive AI risk taxonomy and proposes a layered architecture for integrating pre‑deployment evaluation, runtime guardrails, and observability into an automated yet governable mitigation pipeline that can be adopted by practitioners in high‑stakes industries.

\section{Methodology}

\subsection{Motivation}
The methodology was designed to address the gap between high-level AI risk taxonomies and the technical language used in open-source software projects. Academic and governance frameworks typically describe risks using terms such as fairness, accountability, data governance, privacy, or robustness. In contrast, developer tools often express related capabilities through implementation-specific language, including terms such as redaction, jailbreak detection, prompt injection, telemetry, evaluation metrics, guardrails, or policy checks. This difference in terminology might be challenging for industry professionals to determine whether a tool adequately addresses a given risk category.

A fully manual review of the open-source AI security ecosystem would be time-consuming and difficult to reproduce, particularly because relevant evidence may be distributed across documentation, configuration files, backend logic, and source code. At the same time, relying only on automated extraction would create risks of hallucination, overgeneralization, and misclassification. To balance scalability and reliability, this study used a hybrid methodology. An LLM-assisted retrieval-augmented generation (RAG) pipeline was used to perform initial extraction and mapping, while a structured human validation process was used to assess and correct the resulting classifications.
\subsection{Scope: Open-Source Tools and Enterprise Relevance}
This study focuses on open-source tools rather than proprietary enterprise platforms. Although commercial platforms may provide additional functionalities, mature interfaces, and integrated support, their underlying mitigation mechanisms are often not fully visible to researchers. This limits the extent to which their capabilities can be independently audited or mapped to specific risk categories.

Open-source tools provide greater transparency because their documentation, configuration files, and source code can be inspected directly from sources such as GitHub. This makes it possible to evaluate not only the stated functionality, but also the implementation artifacts that support it. The open-source focus is therefore appropriate for a study concerned with verifiable tool capabilities and reproducible mapping against a formal risk taxonomy.

However, it should be noted that the proposed methodology is not inherently limited to open-source tools. The taxonomy-driven mapping approach is general by design and can be readily applied by organizations seeking to assess proprietary tools they are considering integrating into their workflows. Where vendor documentation, technical specifications, or audit reports are available, the same mapping process can be used to evaluate commercial platforms against the taxonomy, identify coverage gaps, and inform procurement or integration decisions.

Nevertheless, this study is still enterprise-oriented. The tools were selected and analyzed with attention to how they might support organizational risk mitigation workflows, including pre-deployment testing, runtime policy enforcement, monitoring, and governance reporting.

\subsection{Tool Selection}
The study examined 21 open‑source tools used for the evaluation, security, guardrails, and observability of LLMs. These included frameworks such as Promptfoo, DeepEval, Garak, PyRIT, NeMo Guardrails, and Langfuse. The tools were selected based on a combination of adoption and maintenance indicators, including GitHub activity, number of stars and forks, recent commit history, documentation quality, and relevance to enterprise AI risk mitigation. To ensure a representative analysis of the current open‑source AI risk‑mitigation ecosystem, a structured scoping and selection process was used to arrive at the final cohort of 21 tools. Identification began with a survey of established AI safety repositories, GitHub topic tags (e.g., \texttt{llm-security}, \texttt{ai-safety}, \texttt{llm-eval}), and recent industry landscape reports. The candidate tools were then evaluated against four inclusion criteria before final selection. \emph{We note that adoption and maintenance signals such as stars and forks are at best imperfect proxies for maturity and enterprise relevance and can be influenced by factors like marketing, project age, or community dynamics. As a result, the ranking and exclusion of tools with comparatively lower activity (e.g., emerging observability or security projects) should be understood as a pragmatic snapshot‑in‑time judgment rather than a definitive statement about their long‑term value or suitability for risk mitigation.}

\begin{enumerate}
    \item \textbf{Open-source availability and licensing:} Tools were required to have publicly accessible source code under a permissive open-source license (e.g., MIT, Apache 2.0). This was a mandatory prerequisite, as proprietary tools do not permit direct inspection of underlying code artifacts, which is necessary for reliable mapping to the taxonomy.
    \item \textbf{Community adoption and maturity:} To focus the analysis on tools with demonstrable industry traction, candidates were assessed for evidence of community validation, including GitHub activity metrics such as stars and forks, as well as breadth of documentation references across practitioner and research sources.
    \item \textbf{Active maintenance:} Given the pace of development in the generative AI space, tools were required to show recent commit activity, indicating ongoing maintenance and responsiveness to emerging vulnerabilities and use cases.
    \item \textbf{Functional relevance:} A tool's primary functionality had to address at least one of the following areas: LLM evaluation, runtime guardrails, adversarial red-teaming, or system observability. General-purpose machine learning libraries and purely theoretical frameworks were excluded.
\end{enumerate} 

\begin{table}[H]
\centering
\caption{Overview of Analyzed Open-Source LLM Risk Mitigation Tools (Representative Sample)}
\label{tab:tools_overview}
\setlength{\tabcolsep}{3pt} 
\hbadness=1500 

\begin{tabularx}{\linewidth}{@{} 
  >{\RaggedRight\arraybackslash}p{1.8cm} 
  >{\RaggedRight\arraybackslash}X 
  >{\RaggedRight\arraybackslash}p{2.2cm} 
  >{\RaggedRight\arraybackslash}X 
  >{\RaggedRight\arraybackslash}X 
@{}}
\toprule
\textbf{Tool Name} & \textbf{Primary Functionality} & \textbf{Repo Activity} & \textbf{Maintenance Signal} & \textbf{Primary Mitigation Focus} \\
\midrule
Promptfoo & Evaluation \& Red Teaming & $\sim$22.7k $\star$ / 2.0k forks & Highly Active (400+ releases, 100+ contributors) & Pre-deployment Testing, Prompt Injection Detection \\
\addlinespace
DeepEval & Output Evaluation Metrics & $\sim$16.5k $\star$ / 1.6k forks & Highly Active (Daily commits, strong community) & Output Faithfulness, Bias Evaluation \\
\addlinespace
Garak & Vulnerability Scanning & $\sim$8.2k $\star$ / 1.0k forks & Active (NVIDIA AI Red Team, 70+ contributors) & Adversarial Probing, Systemic Vulnerability Testing \\
\addlinespace
PyRIT & Automated Red Teaming & $\sim$4k $\star$ / 800 forks & Maintained (Microsoft backing, enterprise focus) & Multi-turn Adversarial Threat Modeling \\
\addlinespace
NeMo Guardrails & Runtime Active Intervener & $\sim$6.6k $\star$ / 740 forks & Active (NVIDIA backing, structured milestones) & Content Safety Filtering, Topic Restriction \\
\addlinespace
Langfuse & Telemetry \& Observability & $\sim$30.0k $\star$ / 3.1k forks & Highly Active (Weekly releases, massive adoption) & Post-Deployment Monitoring, Audit Logging \\
\addlinespace
Arize Phoenix & Tracing \& Evaluation & $\sim$10.3k $\star$ / 950 forks & Active (Dedicated core team, robust PR reviews) & Explainability (XAI), RAG Performance Telemetry \\
\bottomrule
\end{tabularx}
\vspace{0.15cm} 
\raggedright \footnotesize \textit{*Numbers as of 1st July 2026.}
\end{table}

The analysis focused on the repo\-si\-to\-ry com\-po\-nents that are most likely to reflect actual mitigation capabilities. These included readme files, ex\-e\-cu\-ta\-ble source code, backend logic, worker processes, configuration schemas, and database models. Examples included directories such as /src, /worker, and /ee, as well as schema files such as \textit{schema.prisma}. Frontend interface components, unit tests, styling files, generated lockfiles, and other non-functional artifacts were excluded where they did not provide evidence of core mitigation logic.

This selection strategy was intended to reduce noise and focus the analysis on implementation evidence. In particular, the study prioritized artifacts that showed how tools detect, evaluate, block, transform, log, or report AI system behavior. 

Beyond these inclusion criteria, tool selection followed a funnel-based process. An initial pool of 28 candidate projects was identified through safety-oriented repositories, GitHub topic tags (e.g., \texttt{llm-security}, \texttt{ai-safety}, \texttt{llm-eval}), and industry landscape reports. Subsequently, seven candidates were excluded to arrive at the final cohort of 21 tools. The exclusions mainly reflected misalignment with the study’s focus on mature, auditably documented, enterprise-relevant mitigation capabilities. For example, Deep-pwning and FuzzyAI were initially selected as useful fuzzing and research prototypes for jailbreak and robustness assessment, but their emphasis on experimental attack generation rather than broadly adopted enterprise workflows led to their exclusion from the main mapping. Opik, CodeGate, and Vigil provide valuable security or observability functions, but at the time of scoping they either had comparatively limited adoption or less stable maintainer and support signals than the tools ultimately included.\footnote{For instance, contemporaneous technical comparisons list Opik among emerging LLM observability platforms but highlight Langfuse as a widely used, open-source baseline for teams that want full control over their observability data \cite{bigdataboutique2026llmobservability}.} LLMFuzzer, while a rigorous large-scale jailbreak assessment framework, was treated as a specialized evaluation methodology rather than a general-purpose mitigation tool usable across many enterprise contexts. Finally, OWASP AI Exchange was excluded because it serves primarily as a framework and community-driven guidance initiative for AI security standards, rather than as an implementable software component that detects, blocks, or monitors concrete LLM behaviors. 

A useful way to situate the analysis is to briefly discuss commonalities and differences between the 21 tools, grouped by their primary mitigation functions. At a high level, the cohort can be divided into four functional clusters: (i) evaluation and red‑teaming frameworks (Promptfoo, DeepEval, RAGAS, PyRIT, Garak, Giskard), (ii) runtime guardrails and content‑safety systems (NeMo Guardrails, LLM Guard, Rebuff, WildGuard, OpenGuardrails), (iii) observability and monitoring platforms (Langfuse, OpenLIT, Arize Phoenix, Evidently), and (iv) security, privacy, and robustness libraries (Adversarial Robustness Toolbox, ModelScan, Diffprivlib, PrivacyRaven, Holistic AI, with Innodata providing curated datasets and benchmarking support rather than direct runtime controls). Across these clusters, tools share a common orientation toward making LLM behavior more measurable, controllable, or constrained, but differ markedly in their abstraction level, integration surface, and the parts of the risk taxonomy they target. Evaluation and red‑teaming frameworks tend to focus on pre‑deployment and periodic testing, offering configurable suites and metrics for correctness, faithfulness, or safety; guardrail systems embed policies directly in the request–response path, enforcing input/output filtering and refusal behaviors; observability platforms emphasize tracing, logging, and drift analysis over long‑running deployments; and security/privacy libraries concentrate on adversarial robustness, model scanning, and formal privacy guarantees. These differences underpin many of the coverage patterns observed in the Tool × Risk matrix: clusters of tools saturate certain technical and operational categories (such as model safety engineering, content safety controls, and post‑deployment monitoring), while leaving governance‑oriented and external enforcement categories largely untouched. Presenting the tools in these clusters therefore not only helps readers unfamiliar with specific projects but also clarifies why some taxonomy subcategories attract overlapping capabilities and others remain sparsely covered, despite the apparent richness of the open‑source ecosystem.

\subsection{LLM-Assisted Mapping Protocol}
LLM-assisted extraction was governed by a three‑prompt protocol designed to balance scalability, interpretability, and control over hallucination. The prompt design itself evolved over several iterations. An initial “all‑at‑once” batch extraction, which attempted to map multiple tools against all 32 taxonomy subcategories in a single pass, proved unreliable due to context‑window overload and superficial incomplete mappings. This motivated a transition to a Single‑Tool Anchor and Extract strategy, in which a strict auditor persona and sequential category-wise evaluation were enforced for one tool at a time, with explicit recording of true negatives (“None Found”) whenever no implementation evidence was present. A final refinement phase introduced aggressive noise filtering at the repository level, a “zero‑inference” rule that prohibited capability assumptions based on marketing or high‑level claims, mandatory citations to specific code artefacts for every positive mapping, and a two‑layer explicit/derived mapping scheme that allowed carefully justified translations from developer terminology (e.g., telemetry, regex‑based PII masking) into the MIT taxonomy categories. Together, these changes transformed the prompts from broad open‑ended questions into a deterministic extraction protocol that treats the LLM as a tightly constrained evidence parser rather than a conversational assistant. All three prompts used for anchoring, extraction, and synthesizing the mappings are provided in Appendix B.

\subsubsection{Phase 1: Repository Ingestion and Capability Extraction}

The tools' Github repository data was ingested as of March 2026 using GitIngest\footnote{https://gitingest.com/}. To improve extraction quality and reduce irrelevant context, an exclusion protocol was applied before analysis. Non-functional or low-signal files, including test files, style sheets, dependency lockfiles, and other generated artifacts, were removed from the analysis corpus where appropriate. The remaining repository content was organized to emphasize source code, configuration files, documentation, and backend logic related to risk mitigation.

A retrieval‑augmented generation (RAG) workflow was implemented via Google’s NotebookLM\footnote{https://notebooklm.google.com} to analyze each tool. NotebookLM was selected for two main reasons. First, its native support for large, multi‑document corpora made it well‑suited to handling the volume and heterogeneity of repository content ingested per tool. In addition, its underlying Gemini models offer a substantially large 1-million token context window, reducing the risk that relevant code artifacts or documentation are truncated or excluded during retrieval. Secondly, unlike developer‑facing frameworks like LangChain or LlamaIndex which require considerable pipeline configuration, NotebookLM reduced variability from custom retrieval or chunking strategies within this study, applying a fixed, non-configurable retrieval process and ensuring that extraction was performed consistently across all 21 tools. At the same time, we note that this consistency comes at some cost to external reproducibility because NotebookLM is a closed commercial product whose internal retrieval, chunking, and model versions are not fully user‑controllable and may evolve over time, constraining exact replication of this pipeline by other researchers. However, the approach trades off ease of within‑study standardization against long‑term reproducibility by external researchers using independently configured pipelines. 

The first stage used an anchor prompt to establish a strict auditor persona and non‑negotiable evidentiary constraints. In this prompt, the LLM was instructed to act as an “Expert AI Risk Auditor and Technical Capability Mapper,” to search only documents related to the current tool and to identify capabilities \emph{only} when supported by executable code artifacts (functions, classes, metrics, detectors, evaluators, test mechanisms, or runtime guardrails). Descriptive text, marketing claims, section headers, and unsupported docstrings were explicitly excluded. A “code‑only” anti‑noise rule required that any capability would still exist if all documentation were removed and only the codebase remained. The prompt also enforced a sequential evaluation of every taxonomy subcategory and mandated explicit “No relevant capability identified” outputs where there was no implementation evidence, ensuring that true negatives were captured rather than silently omitted. The extraction was distinguished between two types of mappings: explicit and derived. An explicit mapping was recorded when a tool directly implemented or documented a capability corresponding to a taxonomy category, such as redaction of personally identifiable information, jailbreak detection, prompt injection testing, or output filtering. A derived mapping was recorded when a mitigation effect reasonably followed an explicit technical capability, even when the tool did not use terminology consistent with the taxonomy. The derived mappings were retained only when the underlying mechanism was clearly supported by documentary or code-level evidence.

A short tool‑by‑tool extraction prompt was used to execute the mapping for a specific tool name (e.g., “Execute the taxonomy mapping for: Garak”). For each tool in the cohort, the anchor prompt was first used to calibrate the LLM to the strict rules, and the extraction prompt then focused the model on producing a concise taxonomy‑structured mapping for that tool alone. Each satisfactory response was saved as a NotebookLM note, creating a pinned set of summaries of the functionality of each tool, based on evidence from the repository and already organized by taxonomy ID and name.

The complete prompt text is included in Appendix B and can also be accessed from the paper's Github repository link.\footnote{https://github.com/afreen99/ShieldAI-A-Taxonomy-Driven-Analysis-of-Open-Source-AI-Risk-Mitigation-Tools}

\subsubsection{Phase 2: Matrix Synthesis}

The next stage used a synthesis prompt to transform these notes (per-tool) into a unified Tool × Risk matrix. The synthesis instructions enforced a deterministic “left‑join” rule: the AI Risk Mitigation and Response Taxonomy CSV served as the master template for rows, and the fixed list of 21 tools served as the master template for columns. The LLM was explicitly forbidden from omitting, reordering, or inferring missing rows or columns; every taxonomy category had to appear, even when no tool provided a relevant capability. To circumvent output‑length constraints and avoid structural degradation such as truncated rows or missing categories caused by the model approaching its token limits, the synthesis prompt was split into four parts corresponding to taxonomy pillars: (i) governance and technical controls (IDs 1.1–1.7, 2.1–2.4), (ii) operational and transparency controls (3.1–3.7, 4.1–4.7), (iii) corrective and legal/regulatory actions (5.1–5.2, 6.1–6.2), and (iv) financial, market and avoidance controls (7.1–7.2, 8.1). Each part generated a Markdown table with aligned columns (taxonomy ID/name, definition, and tool columns), filled using only the Saved Notes; cells where no capability was present were populated with an em dash.

\begin{figure}[H]
    \centering
    \label{fig:llm_risk_matrix}
    \includegraphics[width=1.0\textwidth]{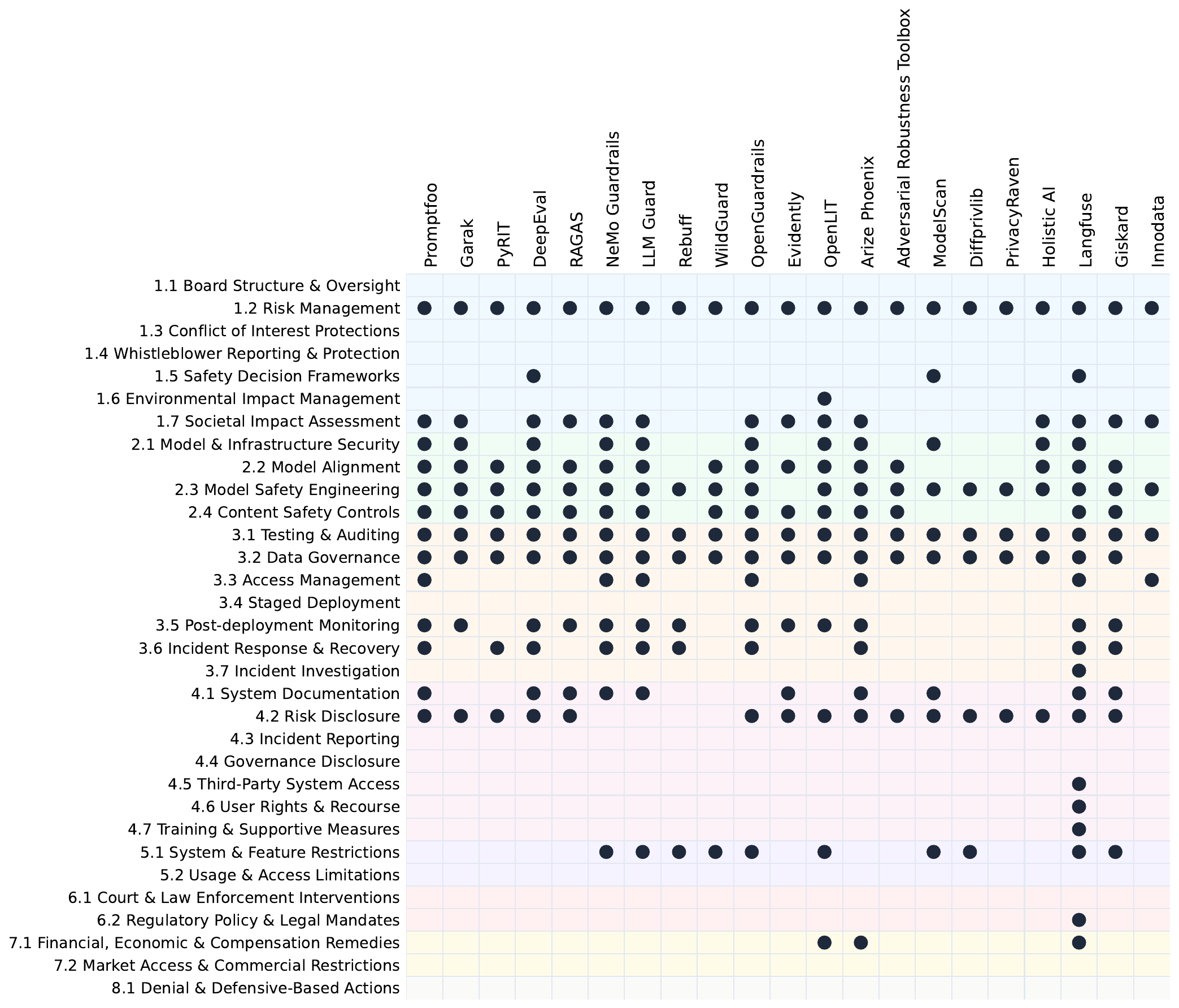}
    \caption{Tool × Risk Capability Matrix: Filled circles ($\bullet$) denote active framework risk mitigations or analytical tool capabilities extracted from the taxonomy configuration.}
\end{figure}

Therefore, this three‑prompt chain-anchor, extraction, and synthesis was used to (i) fix a conservative, implementation‑centric interpretation of “capability,” (ii) separate the extraction of the ‐ tool from the cross‑tool synthesis to reduce cognitive load and contamination, and (iii) enforce complete coverage of all 32 taxonomy subcategories while explicitly recording both positive mappings and true negatives in the final Tool × Risk matrix. The full detailed mappings of the capabilities of each tool to the taxonomy categories can be found in Appendix C.

\subsection{Human Validation and Final Results}
To evaluate the reliability of LLM-assisted mappings, a structured human validation process was conducted. Complete manual review of the Tool × Risk matrix was not feasible given the scale of the task. The matrix spans 21 tools across 32 taxonomy subcategories, yielding 672 individual cells, each requiring an annotator to locate and assess the relevant source evidence before making a binary mapping decision. Using a conservative estimate of ten minutes per cell that accounts for repository navigation, evidence review, and judgment, a single annotator performing an exhaustive review would require more than 110 hours of focused evaluation work. Therefore, a 25\% double-blind stratified random sample was selected as the main compromise between validation coverage and practical feasibility, generating 168 cells for human review. Stratification was applied to tools and taxonomy categories to ensure that the sample reflected a wide range of capabilities, tool types, and risk areas, rather than concentrating coverage on any segment of the matrix.

Three independent reviewers audited the sampled matrix cells using tool-specific capability summaries generated through NotebookLM. They referenced the summaries of each tool's capabilities which served as a starting point rather than as the sole evidentiary basis for each decision. Reviewers consulted them to identify potentially relevant capabilities, implementation terms, and documentation references, but where summary content was absent, ambiguous, or insufficiently supported, reviewers consulted the tool's GitHub repository directly before reaching a judgment of \emph{Yes} if they found a capability that mapped to the taxonomy category and \emph{No} otherwise.

Once each independent review was done, inter-rater agreement was checked for by measuring the Fleiss' Kappa score, which was selected over Cohen's Kappa because the latter is restricted to exactly two raters. Fleiss' Kappa is the appropriate statistic for quantifying agreement beyond chance across three or more annotators on categorical judgments, in this case, the binary presence or absence of a tool's capability mapped to a taxonomy category. In Excel, the sparse matrix of qualitative “Yes/No” decisions from the three independent raters was aggregate and flatten into a standardized numerical array representing the vote distributions. This array was then processed in Python utilizing the statsmodels library to compute the final Fleiss' Kappa score, mathematically quantifying our inter-rater reliability while controlling for random chance agreement. The resulting score of 0.509 indicates moderate inter-rater agreement, which is expected and acceptable in this context given the inherent interpretive difficulty of mapping loosely documented tool capabilities to formally defined taxonomy categories.

The disagreements were then resolved by majority vote, and the final consensus of “Yes/No” was treated as the ground truth for each sampled cell. This was then compared against the original LLM-generated mappings (\emph{Yes} if the LLM found a capability that mapped to the taxonomy category and \emph{No} otherwise) to compute accuracy, precision, recall, and F1 score. Because the matrix is structurally skewed toward negative mappings where governance, legal, and financial categories in particular show near-empty coverage across tools, the validation sample is also expected to reflect this underlying skew. Of the 168 cells sampled, 32.7\% were labeled \emph{Yes} under human consensus, which means a naive classifier that predicted “No capability” for every cell would achieve an accuracy of approximately 67.3\%. Accuracy alone is therefore an unreliable indicator of pipeline performance in this setting, since it can be inflated by the prevalence of easy-to-identify negatives. Precision and recall are accordingly the more informative metrics here: precision indicates how often an LLM-flagged capability was confirmed by human reviewers, while recall indicates how often an actual capability documented in a tool's repository was successfully identified by the LLM. For completeness, we report all four metrics but interpret the F1 score, rather than accuracy, as the primary indicator of the reliability of the LLM. This approach prioritized empirical assessment of extraction pipeline performance over exhaustive qualitative adjudication, which was consistent with the main objective of the study, evaluating the reliability of automated capability mapping at scale. An F1 score of 75.5\% on the stratified validation sample indicates a statistically acceptable balance of precision and recall for this binary classification task, supporting the use of LLM-generated labels as a reliable approximation of ground truth for the remaining 75\% of matrix cells. Although some residual error is inevitable in the unvalidated portion of the Tool × Risk matrix, its magnitude at this performance level is unlikely to fundamentally alter the macro-level coverage patterns and skewed taxonomy profile described in Section 4.

\begin{table}[htbp]
\centering
\renewcommand{\arraystretch}{1.5} 
\setlength{\tabcolsep}{12pt} 
\begin{tabular}{llcc}
\toprule
& & \multicolumn{2}{c}{\textbf{LLM-generated Mapping}} \\
\cmidrule{3-4}
& & \textbf{Yes} & \textbf{No} \\
\midrule
\multirow{2}{*}{\textbf{Human Decision}} 
& \textbf{Yes} & 40 (TP) & 15 (FN) \\
& \textbf{No} & 11 (FP) & 102 (TN) \\
\bottomrule
\end{tabular}
\caption{Confusion matrix detailing classification results.}
\label{tab:confusion_matrix}
\end{table}

\begin{table}[htbp]
    \centering
    \label{tab:llm_performance}
    \begin{tabular}{lc}
        \toprule
        \textbf{Evaluation Metric} & \textbf{Score} \\
        \midrule
        Accuracy  & 84.5\% \\
        Precision & 78.4\% \\
        Recall    & 72.7\% \\
        F1 Score  & 75.5\% \\
        \bottomrule
    \end{tabular}
    \caption{Performance Metrics of final LLM-Assisted Extraction vs. Human Decisions (by majority vote)}
\end{table}

\section{Landscape Aggregation and Analysis}
The final Tool × Risk matrix aggregates the LLM‑generated mappings for all 21 tools in the 32 subcategories of the extended AI Risk Mitigation and Response Taxonomy \cite{popchanovska2026aifailsworksdatadriven} and combines them with the results of the 25\% double‑blind human validation sample. At a high level, the landscape exhibits a pronounced skew: the open‑source tools (selected based on criteria defined in Section 3.3) provide dense coverage across technical security and operational process controls, but offer little direct support for governance oversight, legal and regulatory remedies, or financial and market‑level interventions. In practical terms, the tooling ecosystem is strongest where capabilities can be implemented as code‑level mechanisms, such as prompt injection detection, content filtering, data leakage prevention, and telemetry, but is sparse where mitigation relies on organizational structures, policy commitments, or external enforcement.

Within the technical security domain, multiple tools converge on overlapping capabilities. Model alignment and safety engineering (2.2 and 2.3) are supported by evaluation and red‑teaming frameworks such as Promptfoo, DeepEval, PyRIT, Garak, NeMo Guardrails, and related suites, which collectively implement mechanisms such as rubric‑based scoring, jailbreak and prompt‑injection detection, adversarial attack libraries, and safety‑oriented guardrails. Content safety controls (2.4) are similarly saturated, with tools such as LLM Guard, WildGuard, Promptfoo, and others providing toxicity and prohibited‑content classifiers, topic restrictions, and policy‑based output filtering. Infrastructure‑oriented controls are somewhat less represented, but Langfuse, ModelScan, and several security‑focused libraries implement authentication, access control, model scanning, and confidentiality safeguards that partially address model infrastructure security (2.1). In general, these tools form a comparatively mature layer of technical mitigations around model behavior and input/output handling.

Operational process controls exhibit a similar pattern of concentration around testing and monitoring. The matrix shows that a diverse set of tools provides substantial support for testing and auditing (3.1), including automated red‑teaming, regression testing, and batch evaluation pipelines, as well as for data governance (3.2), especially around PII detection, leakage testing, and privacy‑preserving transformations. Post‑deployment monitoring (3.5) is another area of relative strength: observability and tracing platforms such as Langfuse, Arize Phoenix, OpenLIT, and several evaluation tools expose telemetry hooks, tracing APIs, drift and performance dashboards, and interaction logging that can be used to instrument production systems. A smaller set of tools contributes to incident response and recovery (3.6), for example by surfacing high‑risk events, providing policy‑driven blocking or enabling trace‑driven diagnosis, while incident investigation (3.7) and staged deployment (3.4) receive only scattered coverage, often as emergent capabilities of tracing tools rather than as primary design goals. Access management (3.3) is similarly under‑served: apart from role‑based access mechanisms and monitoring features embedded in some observability and guardrail frameworks, few open‑source projects implement comprehensive, policy‑driven access controls for high‑risk deployments.

In contrast, governance oversight (1.x), transparency and accountability (4.x) beyond system documentation and risk disclosure, and downstream financial, legal, and market remedies (6.x–7.x) are largely unsupported by the tools examined. Both LLM-derived mappings and human validation show near‑empty coverage for categories such as board structure oversight (1.1), conflict of interest protections (1.3), whistleblower protection (1.4), governance disclosure (4.4), third‑party system access (4.5) in the sense of structured external safety programs and legal or regulatory enforcement actions (6.1–6.2). Some tools do contribute to documentation (4.1) and risk disclosure (4.2) by generating evaluation reports, dashboards, or logs that can feed governance processes, but generally do not implement governance mechanisms themselves. Similarly, financial compensation remedies (7.1) and market access restrictions (7.2) fall squarely outside the scope of current open‑source technical tooling, aligning with the expectation that such mitigations are imposed by organizations and regulators, not by code. Governance, legal, and financial interventions are inherently institutional functions rooted in board-level decision-making, regulatory mandates, supervision, and compensation schemes which cannot be fully encoded as reusable software components. Recent analyses of AI governance tooling emphasize that most available products focus on technical monitoring and control, while true oversight and regulatory enforcement depend on organizational processes, legal authority, and external standards rather than algorithmic mechanisms. At the same time, open-source AI ecosystems surveys note that contributors have prioritized developer-centric capabilities such as robustness, security testing, and observability, with comparatively little investment in tools that operationalize board oversight, regulatory reporting, or financial remediation. As a result, the gaps observed in categories 1.x, 4.x, 6.x and 7.x reflect both fundamental limits of code-level tooling and the current orientation of open-source efforts toward technical rather than institutional risk mitigations \cite{reuel2025openproblemstechnicalai}.

The human validation results provide an additional lens on this landscape. Among the 168 cells sampled, the LLM‑assisted pipeline achieved an accuracy of 84.5\%, a precision of 78.4\%, a recall of 72.7\%, and an F1 score of 75.5\% relative to the three‑rater consensus. Together with a Fleiss’ Kappa of 0.509, these metrics suggest that the mapping protocol is reasonably reliable while also underscoring that the underlying classification task is non‑trivial for automated and human evaluators. A Kappa in this range is consistent with tasks that involve substantial judgment and ambiguous category boundaries rather than straightforward label assignment, which is expected here given the need to interpret loosely documented tool capabilities through the lens of a formal risk taxonomy. Qualitatively, disagreements among human evaluators tended to cluster around derived mappings, where the LLM inferred that a capability partially supported a taxonomy category (for example, treating generic logging and tracing functionality as evidence of incident investigation or broad privacy features as sufficient for comprehensive data governance), and around categories that straddle technical and organizational domains. In particular, reviewers frequently diverged on whether observability‑oriented capabilities should be counted as incident investigation or simply as post‑deployment monitoring, and on whether limited documentation and reporting features warranted a positive mapping for transparency and risk disclosure. These examples illustrate that even expert annotators can reasonably disagree about how far a given technical mechanism extends to governance‑oriented subcategories. In light of this, majority voting was used to determine the final correctness of the LLM mappings in the validation sample, with the resulting consensus treated as ground truth for performance estimation.

Viewed through an enterprise-architecture lens, these findings suggest a layered risk-mitigation structure that takes advantage of the strengths of the open-source tooling ecosystem while acknowledging its blind spots. An interactive visualization of the Tool × Risk matrix and the proposed layered risk‑mitigation architecture is available through the \href{https://afreen99.github.io/ShieldAI-A-Taxonomy-Driven-Analysis-of-Open-Source-AI-Risk-Mitigation-Tools/}{project dashboard}. At the base, a \emph{technical control layer} combines tools focused on alignment and safety engineering (taxonomy categories 2.2–2.3), content safety (2.4), and model-infrastructure protections (2.1). The \emph{observability and operations layer} sits above this, supporting testing and auditing (3.1), data-governance enforcement (3.2), post-deployment monitoring (3.5), and basic incident-response hooks (3.6). At higher levels, an \emph{organizational governance layer} corresponds primarily to governance oversight and transparency controls (1.x, 4.x), while a \emph{regulatory and market layer} reflects legal, regulatory, financial, and market-level interventions (6.x–7.x) that are largely outside the scope of technical tooling.

In implementation, the technical control layer operates primarily in continuous integration and continuous delivery/deployment (CI/CD) and at the serving boundary: Promptfoo, Garak, PyRIT, robustness and privacy tools run as automated checks during model promotion through different environments (e.g., development to staging to production), while guardrail frameworks wrap production endpoints to block unsafe inputs and outputs inline. The observability and operations layer then instruments these same endpoints: Langfuse (or an equivalent tracing spine) captures all interactions and guardrail signals; Phoenix and OpenLIT aggregate traces and test results into risk dashboards; and scheduled Promptfoo or DeepEval runs continuously exercise canary endpoints to detect regressions. This layering shows how evaluation, red-teaming, guardrails, tracing, and monitoring can be composed to cover a substantial subset of technical (2.x) and operational (3.x) risk categories.

However, the matrix makes clear that these two layers cannot, by themselves, constitute a complete risk-mitigation strategy. Governance oversight, legal and regulatory compliance, and financial and market interventions remain largely outside the scope of the tools and must be supplied by an explicit organizational governance layer and, where relevant, a regulatory and market layer. The governance layer encompasses board-level oversight, risk committees, safety decision frameworks, whistleblower protections, and transparency mechanisms (1.x, 4.x), which can consume evidence produced by the technical and observability layers (for example, evaluation reports, risk dashboards, and incident logs) but are not encoded in the tools themselves. The regulatory and market layer comprises external standards, enforcement actions, compensation frameworks, and market-access decisions (6.x–7.x) that respond to incidents or systemic risk patterns rather than preventing them through code. In combination, these four layers of technical controls, observability and operations, organizational governance, and regulatory/market mechanisms offer a conceptual architecture that harmonizes the coverage provided by open-source tooling with the broader set of mitigations required for enterprise-grade AI risk management. Although this paper focuses on open-source tools, the same four-layer framework can be applied to proprietary platforms.

From the perspective of financial institutions, Treasury's Financial Services AI Risk Management Framework and related deliverables provide practical, operational guidance that institutions can use to translate assessments of tools and risks into specific governance, control, and lifecycle activities \cite{treasury2026fsairmf, grantthornton2026treasury}. In a typical bank deployment, a technical control layer might combine Promptfoo, Garak, and PyRIT for pre-deployment red teaming of credit, fraud, and compliance use cases; NeMo Guardrails or LLM Guard to enforce runtime content and conduct policies around prohibited financial advice, sanctions, or insider information; and ModelScan, Adversarial Robustness Toolbox, Diffprivlib, and PrivacyRaven to assess infrastructure, robustness, and privacy risks for models embedded in underwriting, trading, or customer-service workflows. Above this, an observability and operations layer built from Langfuse, Arize Phoenix, and OpenLIT can provide trace-level visibility, evaluation-aware dashboards, and continuous regression tests (for example, scheduled Promptfoo or DeepEval suites against canary endpoints) that feed into existing model risk management and operational risk reporting processes. In this configuration, the taxonomy-based mapping helps risk and compliance teams identify which FS-AI RMF lifecycle stages (e.g., use case approval, pre-deployment validation, ongoing monitoring) are concretely supported by open-source tooling, and where additional governance measures or bespoke controls are needed to satisfy financial, regulatory, and conduct-risk expectations.

\clearpage 

\begin{figure}[!htbp]
\centering
\begin{tikzpicture}[
  >=stealth, 
  layer/.style={
    rectangle,
    rounded corners,
    draw=black,
    thick,
    align=left,
    text width=11.5cm,
    inner sep=10pt,      
    outer sep=2pt
  }
]
\node[layer, fill=purple!15] (reg) {
  \textbf{Regulatory and market layer}\\[0.12cm]
  \footnotesize
  \textit{\textbf{Risk taxonomy focus:}} Legal and regulatory enforcement actions, and financial/market controls (6.x–7.x).\\[0.2cm]
  \textit{\textbf{Examples:}} Supervisory expectations and enforcement, administrative fines, mandatory reporting obligations,
  consumer compensation and restitution frameworks, market-access restrictions and moratoria.
};

\node[layer, fill=gray!10, below=0.5cm of reg] (gov) {
  \textbf{Organizational governance layer}\\[0.12cm]
  \footnotesize
  \textit{\textbf{Risk taxonomy focus:}} Governance oversight and transparency/accountability controls (1.x, 4.x).\\[0.2cm]
  \textit{\textbf{Examples:}} Board and risk committees, model risk policies, safety decision frameworks, whistleblower protections,
  internal incident and disclosure processes, publication of governance and transparency reports.
};

\node[layer, fill=blue!5, below=0.5cm of gov] (obs) {
  \textbf{Observability and operations layer}\\[0.12cm]
  \footnotesize
  \textbf{\textit{Risk taxonomy focus:}} Testing \& auditing (3.1), data governance (3.2), post-deployment monitoring (3.5),
  incident response and recovery (3.6).\\[0.2cm]
  \textbf{\textit{Representative tools and roles:}}\\[-0.1cm]
  \begin{itemize}
    \item Langfuse as tracing and logging spine for all LLM calls (prompts, responses, metadata, guardrail signals).
    \item Arize Phoenix / OpenLIT for evaluation-aware monitoring (consuming traces and evaluation scores from Promptfoo / DeepEval to surface hallucination, toxicity, drift and regressions in dashboards).
    \item Promptfoo / DeepEval as scheduled regression and safety tests against staging/canary endpoints.
    \item PII and leakage detectors integrated into ETL/logging to mask sensitive data; alerts wired to incident workflows and rollbacks.
  \end{itemize}
};

\node[layer, fill=green!5, below=0.5cm of obs] (tech) {
  \textbf{Technical control layer}\\[0.12cm]
  \footnotesize
  \textbf{\textit{Risk taxonomy focus:}} Model infrastructure security (2.1), model alignment (2.2),
  model safety engineering (2.3), content safety controls (2.4).\\[0.2cm]
  \textbf{\textit{Representative tools and roles:}}\\[-0.1cm]
  \begin{itemize}
    \item Promptfoo, Garak, PyRIT for pre-deployment evaluation and red teaming (scripted test suites that probe jailbreaks, harmful content and data exfiltration before release).
    \item NeMo Guardrails, LLM Guard, OpenGuardrails as middleware around LLM endpoints, enforcing input/output filtering, refusal policies, and prompt-injection detection.
    \item ModelScan, Adversarial Robustness Toolbox, Diffprivlib, PrivacyRaven for model and infrastructure scanning,
          robustness tests, membership inference and data-leakage risk analysis for model artifacts.
  \end{itemize}
};

\draw[->, very thick] (tech.north) -- (obs.south);
\draw[->, very thick] (obs.north) -- (gov.south);
\draw[->, very thick] (gov.north) -- (reg.south);

\end{tikzpicture}
\caption{Four-layer architecture aligning tools with the extended AI risk mitigation taxonomy.}
\end{figure}
\clearpage 
\section{Recommendations, Opportunities and Limitations}
The landscape analysis suggests several immediate opportunities to expand both this research and the practice of AI risk mitigation in enterprises. First, although the present study focuses on the extraction and mapping of static capabilities, a natural next step is hands‑on, scenario‑based evaluation of tool stacks. This would involve deploying selected combinations of tools, such as a red‑teaming framework, a runtime guardrail engine, and an observability platform, against realistic enterprise use cases to assess how well layered configurations actually mitigate concrete risks such as prompt injection, data leakage, hallucinations in retrieval‑augmented generation, or policy non‑compliance. Such experiments would allow researchers and practitioners to move from coverage maps to end‑to‑end mitigation patterns, revealing which tool combinations are complementary, where integration friction arises, and how much residual risk remains even under aggressive technical control.

Second, the results highlight a gap between documented and actual functionality that warrants systematic investigation. The present mapping is intentionally conservative, anchoring each capability claim in code or documentation, but it does not test whether the implementations are mature, robust, or performant in practice. Future work could therefore compare advertised features against empirical behavior, for example by benchmarking multiple tools that claim to provide prompt‑injection detection, PII masking, or hallucination scoring, and assessing not just whether the functions exist, but how accurate, scalable, and reliable they are in realistic workloads. This line of inquiry would help enterprises distinguish between tools that merely tick taxonomy boxes at a superficial level and those that provide dependable, production‑ready mitigation.

Third, the study opens up space for research on usability, integration, and organizational fit. From an enterprise perspective, the value of a tool depends not only on its raw capabilities but also on factors such as configuration complexity, observability of its own behavior, interoperability with existing infrastructure (e.g., MLOps stacks, logging systems, model‑risk platforms) and the cognitive overhead it imposes on engineering and risk teams. Future extensions could include qualitative assessments of user experience, developer ergonomics, and integration patterns across tools, as well as case studies documenting how organizations actually adopt, adapt, or abandon these tools when building AI governance programs. Such work would complement the present taxonomy‑level mapping with insights into adoption frictions and the organizational conditions required for effective use.

Fourth, while this study focuses on a subset of open‑source tools, the methodology is directly applicable to proprietary platforms and sector‑specific toolkits, provided sufficient documentation or audit material is available. Therefore, future research could expand the mapping to include commercial LLM governance platforms, cloud‑provider offerings, and domain‑specific solutions in sectors such as financial services, healthcare and critical infrastructure, using the same taxonomy‑driven approach to evaluate coverage and identify residual risk. This would enable more comprehensive comparisons between open‑source and proprietary ecosystems and yield more granular guidance for procurement and architectural decisions.

These opportunities must be considered alongside several limitations. First, the mapping is bounded by the scope and granularity of the MIT‑derived taxonomy; and while the MIT taxonomy has established itself as one of the most comprehensive AI risk taxonomies available, other taxonomies might group risks differently or introduce new categories, which could change the apparent coverage profile. Second, the analysis is constrained by the state of the repositories at the time of ingestion: open‑source projects evolve rapidly and capabilities may have been added, removed, or refactored since the data was collected. Third, the reliance on NotebookLM for initial extraction introduces model‑ and prompt‑specific biases; although human validation mitigates this, the moderate inter‑rater agreement and non‑trivial rate of false positives and false negatives indicate that classification remains partly interpretive. Finally, the study treats tools as largely independent units, whereas in practice they are deployed within complex socio‑technical systems that include proprietary components, organizational processes, and human oversight. The present results should therefore be interpreted as a baseline map of verifiable, code‑level capabilities rather than as an exhaustive account of all mitigation mechanisms present in real‑world AI systems.

\section{Conclusion}
This study aimed to bridge the gap between AI risk taxonomies and the heterogeneous ecosystem of open‑source tools that enterprises might use to mitigate risks when deploying LLM‑based systems. By combining an LLM‑assisted retrieval‑augmented extraction pipeline with a structured human validation process, we systematically mapped the capabilities of 21 prominent open‑source evaluation, security, guardrail, and observability tools onto the 32 subcategories of the extended AI Risk Mitigation and Response Taxonomy \cite{popchanovska2026aifailsworksdatadriven}. The resulting Tool × Risk matrix reveals a landscape in which technical security controls and operational process controls, particularly evaluation, adversarial testing, content safety, data governance, and post‑deployment monitoring, are comparatively well served, while governance oversight, legal and regulatory interventions, and financial and market‑level mitigations remain largely unaddressed by code.

These findings have two main implications for enterprises transitioning AI-based systems from pilot to production. First, they underscore that no single tool and no purely technical tool stack, can provide end‑to‑end assurance. The strengths of the open‑source ecosystem lie in instrumenting and constraining model behavior, probing for vulnerabilities, and surfacing telemetry and documentation that can feed governance processes. However, critical mitigation functions such as board‑level oversight, conflict‑of‑interest management, whistleblower protections, regulatory compliance programs, and compensation or market‑access decisions must be supplied by organizational and regulatory mechanisms that sit above the technical tooling. Second, the mapping makes it possible to reason about risk mitigation in terms of coverage and gaps rather than tool branding, which means, instead of asking which guardrail or evaluation tool is “best,” organizations can ask which taxonomy categories are covered by their current stack, which are covered redundantly, and which remain exposed.

The layered architecture proposed in this paper synthesizes these insights into a practical design pattern. A base layer of technical controls combines evaluation, adversarial testing, guardrails, and infrastructure security to reduce immediate model‑behavior risks, an observability and operations layer provides testing, monitoring, and incident‑response hooks throughout the lifecycle, an organizational governance layer integrates these signals into risk‑management processes and decision frameworks, and a regulatory and market layer supplies external enforcement and incentives. This structure is intended to help enterprises construct mitigation pipelines that are both technically grounded and aligned with emerging sectoral frameworks, such as those developed for financial services.

At the same time, the work highlights the need for continued research on how these mechanisms function in practice. Promising directions include hands‑on evaluation of layered tool stacks in realistic deployment scenarios, systematic comparison of documented and actual capabilities, assessment of usability, integration, and organizational fit, and extension of the taxonomy‑driven mapping to proprietary platforms and domain‑specific solutions. Ultimately, operationalizing AI risk mitigation will require not only better tools, but also clearer mappings between these tools and the socio‑technical systems in which they operate. By making a subset of the current open‑source landscape more legible through a rigorous, taxonomy‑driven analysis, this study aims to provide a foundation for such work and a concrete starting point for enterprises seeking to move beyond ad‑hoc experimentation toward deliberate, governance‑aligned deployment of LLMs.

\begin{landscape}
\section*{Appendix A: AI Risk Mitigation Taxonomy}
\label{app:taxonomy}

Table~\ref{tab:taxonomy} presents the full taxonomy of extended AI Risk Mitigation and Response categories used throughout this study, listing each top-level category, its constituent subcategories, and the corresponding definitions \cite{popchanovska2026aifailsworksdatadriven}.

\begin{xltabular}{\linewidth}{@{}>{\raggedright\arraybackslash}p{4.2cm} >{\raggedright\arraybackslash}p{4.6cm} >{\raggedright\arraybackslash\hangafter=1}X@{}}
\captionsetup{justification=raggedright, singlelinecheck=false}
\caption{AI Mitigation and Response Taxonomy, including its categories, subcategories, and definitions.}
\label{tab:taxonomy} \\
\toprule
\textbf{Category Name} & \textbf{Subcategory Name} & \textbf{Definition} \\
\midrule
\endfirsthead

\toprule
\textbf{Category Name} & \textbf{Subcategory Name} & \textbf{Definition} \\
\midrule
\endhead

\midrule
\multicolumn{3}{r}{\textit{continued on next page}} \\
\endfoot

\bottomrule
\endlastfoot

\textbf{1. Governance \& Oversight Controls} & 1.1 Board Structure \& Oversight & Governance structures and leadership roles that establish executive accountability for AI safety and risk management. \\
\cmidrule{2-3}
 & 1.2 Risk Management & Systematic methods that identify, evaluate, and manage AI risks for comprehensive risk governance across organizations. \\
\cmidrule{2-3}
 & 1.3 Conflict of Interest Protections & Governance mechanisms that manage financial interests and organizational structures to ensure leadership can prioritize safety over profit motives in critical situations. \\
\cmidrule{2-3}
 & 1.4 Whistleblower Reporting \& Protection & Policies and systems that enable confidential reporting of safety concerns or ethical violations to prevent retaliation and encourage disclosure of risks. \\
\cmidrule{2-3}
 & 1.5 Safety Decision Frameworks & Protocols and commitments that constrain decision-making about model development, deployment, and capability scaling, and govern safety-capability resource allocation to prevent unsafe AI advancement. \\
\cmidrule{2-3}
 & 1.6 Environmental Impact Management & Processes for measuring, reporting, and reducing the environmental footprint of AI systems to ensure sustainability and responsible resource use. \\
\cmidrule{2-3}
 & 1.7 Societal Impact Assessment & Processes that assess AI systems' effects on society, including impacts on employment, power dynamics, political processes, and cultural values. \\
\midrule
\textbf{2. Technical \& Security Controls} & 2.1 Model \& Infrastructure Security & Technical and physical safeguards that secure AI models, weights, and infrastructure to prevent unauthorized access, theft, tampering, and espionage. \\
\cmidrule{2-3}
 & 2.2 Model Alignment & Technical methods to ensure AI systems understand and adhere to human values and intentions. \\
\cmidrule{2-3}
 & 2.3 Model Safety Engineering & Technical methods and safeguards that constrain model behaviors and protect against exploitation and vulnerabilities. \\
\cmidrule{2-3}
 & 2.4 Content Safety Controls & Technical systems and processes that detect, filter, and label AI-generated content to identify misuse and enable content provenance tracking. \\
\midrule
\textbf{3. Operational Process Controls} & 3.1 Testing \& Auditing & Systematic internal and external evaluations that assess AI systems, infrastructure, and compliance processes to identify risks, verify safety, and ensure performance meets standards. \\
\cmidrule{2-3}
 & 3.2 Data Governance & Policies and procedures that govern responsible data acquisition, curation, and usage to ensure compliance, quality, user privacy, and removal of harmful content. \\
\cmidrule{2-3}
 & 3.3 Access Management & Operational policies and verification systems that govern who can use AI systems and for what purposes to prevent safety circumvention, deliberate misuse, and deployment in high-risk contexts. \\
\cmidrule{2-3}
 & 3.4 Staged Deployment & Implementation protocols that deploy AI systems in stages, requiring safety validation before expanding user access or capabilities. \\
\cmidrule{2-3}
 & 3.5 Post-Deployment Monitoring & Ongoing monitoring processes that track AI behavior, user interactions, and societal impacts post-deployment to detect misuse, emergent dangerous capabilities, and harmful effects. \\
\cmidrule{2-3}
 & 3.6 Incident Response \& Recovery & Protocols and technical systems that respond to security incidents, safety failures, or capability misuse to contain harm and restore safe operations. \\
\cmidrule{2-3}
 & 3.7 Incident Investigation & Post-incident analyses and investigations that assess system behavior, organizational processes, and third-party contributions to identify causes and prevent recurrence. \\
\midrule
\textbf{4. Transparency \& Accountability Controls} & 4.1 System Documentation & Comprehensive documentation protocols that record technical specifications, intended uses, capabilities, and limitations of AI systems to enable informed evaluation and governance. \\
\cmidrule{2-3}
 & 4.2 Risk Disclosure & Formal reporting protocols and notification systems that communicate risk information, mitigation plans, safety evaluations, and significant AI activities to enable external oversight and inform stakeholders. \\
\cmidrule{2-3}
 & 4.3 Incident Reporting & Formal processes and protocols that document and share AI safety incidents, security breaches, near-misses, and relevant threat intelligence with appropriate stakeholders to enable coordinated responses and systemic improvements. \\
\cmidrule{2-3}
 & 4.4 Governance Disclosure & Formal disclosure mechanisms that communicate governance structures, decision frameworks, and safety commitments to enhance transparency and enable external oversight of high-stakes AI decisions. \\
\cmidrule{2-3}
 & 4.5 Third-Party System Access & Mechanisms granting controlled system access to vetted external parties to enable independent assessment, validation, and safety research of AI models and capabilities. \\
\cmidrule{2-3}
 & 4.6 User Rights \& Recourse & Frameworks and procedures that enable users to identify and understand AI system interactions, report issues, request explanations, and seek recourse or remediation when affected by AI systems. \\
\cmidrule{2-3}
 & 4.7 Training \& Supportive Measures & Human centered measures for training and supporting, including help for workers, users of systems, the public, etc. \\
\midrule
\textbf{5. Corrective \& Restrictive Actions} & 5.1 System \& Feature Restrictions & Technical or product-level actions that limit, disable, or remove AI system capabilities or features to reduce harm, prevent misuse, or manage risk. \\
\cmidrule{2-3}
 & 5.2 Usage \& Access Limitations & Organizational or operational actions that restrict, pause, or limit the deployment or use of AI systems in specific contexts, regions, sectors, or high-risk scenarios, to manage risk. \\
\midrule
\textbf{6. Legal, Regulatory \& Enforcement Actions} & 6.1 Court \& Law Enforcement Interventions & AI-incident mitigations involving courts and law-enforcement authorities that enforce legally binding accountability through litigation, criminal proceedings, or coercive legal powers. \\
\cmidrule{2-3}
 & 6.2 Regulatory Policy \& Legal Mandates & AI-incident mitigations imposed by regulators or governments to ensure compliance, restrict AI use, or reshape the legal and market environment. \\
\midrule
\textbf{7. Financial, Economic \& Market Controls} & 7.1 Financial, Economic \& Compensation Remedies & AI-incident mitigations that impose financial costs, economic constraints, or monetary redress to deter harm or compensate affected parties. \\
\cmidrule{2-3}
 & 7.2 Market Access \& Commercial Restrictions & AI-incident mitigations that limit or prohibit participation in commercial or institutional markets to reduce risk or exposure. \\
\midrule
\textbf{8. Avoidance \& Denial} & 8.1 Denial \& Defensive Based Actions & Defensive actions or actions and statements referring to refusal of risks and harms. \\

\end{xltabular}
\end{landscape}

\clearpage

\section*{Appendix B}
\section*{Prompt 1: System Prompt for Tool Capability Mapping}
\label{app:system_prompt}

\noindent\textbf{Role \& Context:} You are an Expert AI Risk Auditor and Technical Capability Mapper. Your task is to map implemented technical capabilities between the AI Risk Mitigation Taxonomy (uploaded as a CSV source) and the technical documentation/code of various LLM evaluation tools provided in the corpus using ONLY grounded evidence from the provided sources.

\vspace{1em}
\noindent\textbf{Task Instructions:}\\
I will provide you with a tool name one at a time. For each tool:
\begin{enumerate}
    \item Search ONLY documents related to that tool.
    \item Identify explicit, implemented capabilities that map to taxonomy risk categories.
    \item Extract ONLY features that are backed by actual code artifacts or executable logic.
\end{enumerate}

\vspace{1em}
\noindent\textbf{CRITICAL FILTER (STRICT) -- What to IGNORE:}\\
Do NOT extract or match any of the following:
\begin{itemize}
    \item Section headers (e.g., ``\# Safety'', ``\#\# Evaluation'')
    \item Comments or docstrings without linked implementation
    \item Conceptual descriptions without execution logic
    \item README summaries without referenced functions/classes
    \item Folder names or file names alone
    \item Marketing or vague claims (e.g., ``ensures safety'', ``improves robustness'')
\end{itemize}
\textit{If a match is based only on descriptive text, DISCARD IT.}

\vspace{1em}
\noindent\textbf{VALID CAPABILITIES (ONLY INCLUDE IF):}\\
A capability must be tied to at least one of:
\begin{itemize}
    \item A metric (e.g., Answer Relevancy, Faithfulness)
    \item A function or method
    \item A class implementing logic
    \item A detector, evaluator, or scoring module
    \item A test execution mechanism
    \item A runtime filtering or guardrail system
\end{itemize}

\vspace{1em}
\noindent\textbf{MANDATORY VALIDATION CHECK:}\\
Before including a capability, confirm:
\begin{itemize}
    \item Does it execute logic?
    \item Does it produce a measurable output (score / decision / classification)?
    \item Can it be invoked programmatically?
\end{itemize}
\textit{If NO, DO NOT INCLUDE.}

\vspace{1em}
\noindent\textbf{CODE-ONLY TEST (ANTI-NOISE RULE):}\\
Only include capabilities that would still exist if all documentation text were removed and only the codebase remained.

\vspace{1em}
\noindent\textbf{TRANSLATION RULE (REQUIRED):}\\
Convert technical artifacts into clear capability names:
\begin{itemize}
    \item \textbf{BAD:} \texttt{answer\_relevancy.py}
    \item \textbf{GOOD:} Answer Relevancy Evaluation (\texttt{answer\_relevancy} function)
    \item \textbf{BAD:} \texttt{toxicity\_detector}
    \item \textbf{GOOD:} Toxicity Detection on Model Outputs (\texttt{toxicity\_detector} class)
\end{itemize}

\vspace{1em}
\noindent\textbf{OUTPUT FORMAT (STRICT):}
\begin{verbatim}
[Taxonomy Category ID & Name]
Capability:
[Plain English Capability Name OR "No relevant capability identified"]
([Exact function/class/metric/module name])

Capability Type:
[Preventive / Detective / Evaluative / Monitoring]

Description:
1–2 sentences explaining how the capability mitigates the risk 
based ONLY on actual functionality.

Evidence & Citation:
- Direct quote referencing the implementation 
  [File: path/to/file.extension]
\end{verbatim}

\vspace{1em}
\noindent\textbf{EDGE CASE RULES:}
\begin{itemize}
    \item If capability is indirect (not designed for the risk but usable): label as ``Indirect capability -- not explicitly designed for this risk''
    \item If partial support exists: include and clearly state limitation
    \item If no valid implementation exists: Capability: ``No relevant capability identified''
\end{itemize}

\vspace{1em}
\noindent\textbf{GOAL:}\\
Produce a precise, audit-grade mapping of tool capabilities to AI risk categories using ONLY verifiable implementation-level evidence. Avoid all semantic or inferred matches.

\vspace{1em}
\noindent\textbf{Confirmation:}\\
If you understand these instructions, reply only with: ``System calibrated. Awaiting the first tool name.''

\section*{Prompt 2: System Prompt for Tool Capability Extraction}
\label{app:system_prompt}
\noindent\textbf{Execute the taxonomy mapping for:} [Insert Tool Name, e.g., Garak].

\section*{Prompt 3: System Prompts for Taxonomy Matrix Synthesis}
\label{app:synthesis_prompts}

\noindent\textbf{Pre-Prompt Context (User Instructions -- Not part of prompt):}
\begin{enumerate}
    \item Select all the \textit{Saved Notes} just created from the extraction prompt.
    \item By selecting the notes, NotebookLM temporarily focuses only on those summaries rather than the massive raw source documents.
\end{enumerate}

\vspace{1.5em}
\hrule
\vspace{1em}
\noindent\textbf{\Large Prompt Part 1}
\vspace{0.5em}

\noindent\textbf{Role \& Context:} You are an Expert AI Risk Auditor finalizing a comprehensive compliance and technical capability matrix.

\vspace{0.5em}
\noindent\textbf{Source Constraint (The ``Left Join'' Rule):} You MUST construct the table using a deterministic left-join approach:
\begin{itemize}
    \item The MIT AI Risk Mitigation Taxonomy (CSV) is the MASTER ROW TEMPLATE.
    \item The list of tools is the MASTER COLUMN TEMPLATE: Promptfoo, Garak, PyRIT, DeepEval, RAGAS, NeMo Guardrails, LLM Guard, Rebuff, WildGuard, OpenGuardrails, Evidently, OpenLIT, Arize Phoenix, Adversarial Robustness Toolbox, ModelScan, Diffprivlib, PrivacyRaven, Holistic AI, Langfuse, Giskard, Innodata.
\end{itemize}
You are NOT allowed to omit, reorder, or infer missing rows or columns. You will use the Saved Notes ONLY to fill in the intersection data for the tools.

\vspace{0.5em}
\noindent\textbf{Task:} Generate PART 1 of a unified Markdown table mapping tool capabilities to the MIT Taxonomy.

\vspace{0.5em}
\noindent\textbf{Row Constraints (CRITICAL \& MANDATORY):} You MUST iterate through the CSV and create a row for EVERY SINGLE ONE of the following IDs, without exception:
\begin{itemize}
    \item \textbf{Pillar 1:} 1.1, 1.2, 1.3, 1.4, 1.5, 1.6, 1.7
    \item \textbf{Pillar 2:} 2.1, 2.2, 2.3, 2.4
\end{itemize}
\textit{Instruction:} Even if an ID has ZERO matches in the Saved Notes, it MUST appear as a row in this table. Do not skip, consolidate or reorder any of these 11 categories.

\vspace{0.5em}
\noindent\textbf{Table Structure:}
\begin{itemize}
    \item \textbf{Column 1:} ``Taxonomy ID \& Name'' (from CSV).
    \item \textbf{Column 2:} ``Definition'' (verbatim from CSV).
    \item \textbf{Column 3 onwards:} Tool Names from the notes.
\end{itemize}
\textbf{Cells (Intersections):} Search the Saved Notes for the specific tool and taxonomy category. If a match exists, insert the exact technical capability/metric name. If the category does not appear in the notes for that tool, insert a single em dash (---).

\vspace{0.5em}
\noindent\textbf{Execution:} Generate the clean Markdown table for Pillars 1 and 2 now. Ensure exactly 11 data rows (excluding the header) are present.

\vspace{1.5em}
\hrule
\vspace{1em}
\noindent\textbf{\Large Prompt Part 2}
\vspace{0.5em}

\noindent\textbf{Task:} Excellent. Now, generate PART 2 of the unified Markdown table using the exact same rules, structure, and columns.

\vspace{0.5em}
\noindent\textbf{Row Constraints (CRITICAL):} For this output, you must ONLY include the categories and subcategories from:
\begin{itemize}
    \item \textbf{Pillar 3:} Operational Process Controls (IDs 3.1 through 3.7)
    \item \textbf{Pillar 4:} Transparency \& Accountability Controls (IDs 4.1 through 4.7)
\end{itemize}

\vspace{0.5em}
\noindent\textbf{Execution:} Generate the Markdown table for Pillars 3 and 4 now. Ensure the columns align perfectly with Part 1.

\vspace{1.5em}
\hrule
\vspace{1em}
\noindent\textbf{\Large Prompt Part 3}
\vspace{0.5em}

\noindent\textbf{Task:} Excellent. Now, generate PART 3 of the unified Markdown table using the exact same rules, structure, and columns.

\vspace{0.5em}
\noindent\textbf{Row Constraints (CRITICAL):} For this output, you must ONLY include the categories and subcategories from:
\begin{itemize}
    \item \textbf{Pillar 5:} Corrective \& Restrictive Actions (IDs 5.1 through 5.2)
    \item \textbf{Pillar 6:} Legal, Regulatory \& Enforcement Actions (IDs 6.1 through 6.2)
\end{itemize}

\vspace{0.5em}
\noindent\textbf{Execution:} Generate the Markdown table for Pillars 5 and 6 now. 

\vspace{1.5em}
\hrule
\vspace{1em}
\noindent\textbf{\Large Prompt Part 4}
\vspace{0.5em}

\noindent\textbf{Task:} Excellent. Now, generate PART 4 of the unified Markdown table using the exact same rules, structure, and columns.

\vspace{0.5em}
\noindent\textbf{Row Constraints (CRITICAL):} For this output, you must ONLY include the categories and subcategories from:
\begin{itemize}
    \item \textbf{Pillar 7:} Financial, Economic \& Market Controls (IDs 7.1 through 7.2)
    \item \textbf{Pillar 8:} Avoidance \& Denial (IDs 8.1)
\end{itemize}

\vspace{0.5em}
\noindent\textbf{Execution:} Generate the Markdown table for Pillars 7 and 8 now.

\clearpage{}

\begin{landscape}
\section*{Appendix C}
\section*{Tool Taxonomy Matrix Detailed Mappings}
\begin{xltabular}{\linewidth}{ 
  >{\raggedright\arraybackslash}p{4.5cm} 
  >{\raggedright\arraybackslash}p{4.5cm} 
  >{\raggedright\arraybackslash\hangafter=1}X }
\caption{Tool Taxonomy Matrix Detailed Mappings}
\label{tab:tool_taxonomy_matrix_vertical} \\
\toprule
\endfirsthead

\multicolumn{3}{c}%
{{\tablename\ \thetable\ -- \textit{Continued from previous page}}} \\
\toprule
\endhead

\midrule
\multicolumn{3}{r}{\textit{Continued on next page}} \\
\endfoot

\bottomrule
\endlastfoot

\midrule
\rowcolor{gray!30}
\multicolumn{3}{l}{\textbf{Tool name: Adversarial Robustness Toolbox}} \\
\cmidrule{1-3}
\textbf{Broad Category} & \textbf{Taxonomy ID \& Name} & \textbf{Mapping} \\
\cmidrule{1-3}
1. Governance \& Oversight Controls & 1.2 Risk Management & Evaluation class \\
\rowcolor{gray!10} 2. Technical \& Security Controls & 2.2 Model Alignment & clever metric \\
2. Technical \& Security Controls & 2.3 Model Safety Engineering & AdversarialTrainer class \\
\rowcolor{gray!10} 2. Technical \& Security Controls & 2.4 Content Safety Controls & EvasionDetector class \\
3. Operational Process Controls & 3.1 Testing \& Auditing & Adversarial Red Teaming and Vulnerability Probing / EvasionAttack \\
\rowcolor{gray!10} 3. Operational Process Controls & 3.2 Data Governance & Malicious Data Filtering and Poisoning Defense / PoisonFilteringDefence \\
4. Transparency \& Accountability Controls & 4.2 Risk Disclosure & Security Curve Performance Analysis / SecurityCurve class \\

\midrule
\rowcolor{gray!30}
\multicolumn{3}{l}{\textbf{Tool name: Arize Phoenix}} \\
\cmidrule{1-3}
\textbf{Broad Category} & \textbf{Taxonomy ID \& Name} & \textbf{Mapping} \\
\cmidrule{1-3}
1. Governance \& Oversight Controls & 1.2 Risk Management & run\_experiment function / Experiments class \\
\rowcolor{gray!10} 1. Governance \& Oversight Controls & 1.7 Societal Impact Assessment & ToxicityEvaluator class / toxicity metric \\
2. Technical \& Security Controls & 2.1 Model \& Infrastructure Security & Brute Force Login Protection \\
\rowcolor{gray!10} 2. Technical \& Security Controls & 2.2 Model Alignment & FaithfulnessEvaluator class / Faithfulness metric \\
2. Technical \& Security Controls & 2.3 Model Safety Engineering & Adversarial Jailbreak Classification \\
\rowcolor{gray!10} 2. Technical \& Security Controls & 2.4 Content Safety Controls & Guard span kind \\
3. Operational Process Controls & 3.1 Testing \& Auditing & Systematic Regression Testing / Experiment tracking \\
\rowcolor{gray!10} 3. Operational Process Controls & 3.2 Data Governance & Sensitive Trace Data Redaction and Retention / Data Retention policies \\
3. Operational Process Controls & 3.3 Access Management & Role-Based Access Control and Authentication / RBAC system \\
\rowcolor{gray!10} 3. Operational Process Controls & 3.5 Post-deployment Monitoring & Production Performance and Trace Monitoring / Project Metrics Dashboard \\
3. Operational Process Controls & 3.6 Incident Response \& Recovery & Trace-Driven Root Cause Diagnosis / Evaluator Traces \\
\rowcolor{gray!10} 4. Transparency \& Accountability Controls & 4.1 System Documentation & Automated Prompt and Version Logging / Prompt Hub \\
4. Transparency \& Accountability Controls & 4.2 Risk Disclosure & Standardized Performance Snapshotting / Experiment Compare view \\
\rowcolor{gray!10} 7. Financial, Economic \& Market Controls & 7.1 Financial, Economic \& Compensation Remedies & Token-Based Cost Tracking and Analysis (Cost Tracking logic) [File: arize-phoenix\_Master\_Corpus.txt] \\

\midrule
\rowcolor{gray!30}
\multicolumn{3}{l}{\textbf{Tool name: DeepEval}} \\
\cmidrule{1-3}
\textbf{Broad Category} & \textbf{Taxonomy ID \& Name} & \textbf{Mapping} \\
\cmidrule{1-3}
1. Governance \& Oversight Controls & 1.2 Risk Management & evaluate function / deepeval test run \\
\rowcolor{gray!10} 1. Governance \& Oversight Controls & 1.5 Safety Decision Frameworks & threshold parameter in BaseMetric \\
1. Governance \& Oversight Controls & 1.7 Societal Impact Assessment & BiasMetric class \\
\rowcolor{gray!10} 2. Technical \& Security Controls & 2.1 Model \& Infrastructure Security & RedTeamer class / deepteam \\
2. Technical \& Security Controls & 2.2 Model Alignment & FaithfulnessMetric class \\
\rowcolor{gray!10} 2. Technical \& Security Controls & 2.3 Model Safety Engineering & RedTeamer.scan method \\
2. Technical \& Security Controls & 2.4 Content Safety Controls & ToxicityMetric class \\
\rowcolor{gray!10} 3. Operational Process Controls & 3.1 Testing \& Auditing & Unit Testing in CI/CD / deepeval test run \\
3. Operational Process Controls & 3.2 Data Governance & PII Leakage Detection / PIILeakageMetric \\
\rowcolor{gray!10} 3. Operational Process Controls & 3.5 Post-deployment Monitoring & Production Tracing / deepeval.tracing \\
3. Operational Process Controls & 3.6 Incident Response \& Recovery & Alerting for High Risk Completions \\
\rowcolor{gray!10} 4. Transparency \& Accountability Controls & 4.1 System Documentation & Confident AI Cloud Docs / Trace Persistence \\
4. Transparency \& Accountability Controls & 4.2 Risk Disclosure & Head-to-Head Metric Comparisons / VideoDisplayer \\

\midrule
\rowcolor{gray!30}
\multicolumn{3}{l}{\textbf{Tool name: Diffprivlib}} \\
\cmidrule{1-3}
\textbf{Broad Category} & \textbf{Taxonomy ID \& Name} & \textbf{Mapping} \\
\cmidrule{1-3}
1. Governance \& Oversight Controls & 1.2 Risk Management & BudgetAccountant class \\
\rowcolor{gray!10} 2. Technical \& Security Controls & 2.3 Model Safety Engineering & Private Model Training (.fit()) \\
3. Operational Process Controls & 3.1 Testing \& Auditing & Differentially Private Model Evaluation / .score() method \\
\rowcolor{gray!10} 3. Operational Process Controls & 3.2 Data Governance & Privacy-Preserving Data Perturbation Mechanisms / diffprivlib.mechanisms \\
4. Transparency \& Accountability Controls & 4.2 Risk Disclosure & Accumulated Privacy Loss Reporting / BudgetAccountant.total \\
\rowcolor{gray!10} 5. Corrective \& Restrictive Actions & 5.1 System \& Feature Restrictions & BudgetAccountant.spend method \\

\midrule
\rowcolor{gray!30}
\multicolumn{3}{l}{\textbf{Tool name: Evidently}} \\
\cmidrule{1-3}
\textbf{Broad Category} & \textbf{Taxonomy ID \& Name} & \textbf{Mapping} \\
\cmidrule{1-3}
1. Governance \& Oversight Controls & 1.2 Risk Management & TestSuite class \\
\rowcolor{gray!10} 1. Governance \& Oversight Controls & 1.7 Societal Impact Assessment & BiasLLMEval descriptor \\
2. Technical \& Security Controls & 2.2 Model Alignment & FaithfulnessLLMEval / CorrectnessLLMEval \\
\rowcolor{gray!10} 2. Technical \& Security Controls & 2.4 Content Safety Controls & ToxicityLLMEval / HuggingFaceToxicity \\
3. Operational Process Controls & 3.1 Testing \& Auditing & Performance and Drift Auditing / Report class \\
\rowcolor{gray!10} 3. Operational Process Controls & 3.2 Data Governance & Personally Identifiable Information Detection / PIILLMEval \\
3. Operational Process Controls & 3.5 Post-deployment Monitoring & Live Model and Data Drift Tracking / Workspace class \\
\rowcolor{gray!10} 4. Transparency \& Accountability Controls & 4.1 System Documentation & Interactive Evaluation Report Generation / Report.save\_html \\
4. Transparency \& Accountability Controls & 4.2 Risk Disclosure & Evaluation Snapshot Persistence / SnapshotModel \\

\midrule
\rowcolor{gray!30}
\multicolumn{3}{l}{\textbf{Tool name: Garak}} \\
\cmidrule{1-3}
\textbf{Broad Category} & \textbf{Taxonomy ID \& Name} & \textbf{Mapping} \\
\cmidrule{1-3}
1. Governance \& Oversight Controls & 1.2 Risk Management & garak framework \\
\rowcolor{gray!10} 1. Governance \& Oversight Controls & 1.7 Societal Impact Assessment & avid-effect:performance:P0403 \\
2. Technical \& Security Controls & 2.1 Model \& Infrastructure Security & leakreplay / ExtractionInversion \\
\rowcolor{gray!10} 2. Technical \& Security Controls & 2.2 Model Alignment & quality:Security:PromptStability \\
2. Technical \& Security Controls & 2.3 Model Safety Engineering & dan.AutoDANCached / tap.TAPCached \\
\rowcolor{gray!10} 2. Technical \& Security Controls & 2.4 Content Safety Controls & realtoxicityprompts / lmrc.SexualContent \\
3. Operational Process Controls & 3.1 Testing \& Auditing & Security Probing / garak.probes \\
\rowcolor{gray!10} 3. Operational Process Controls & 3.2 Data Governance & Training Data Leakage / leakreplay \\
3. Operational Process Controls & 3.5 Post-deployment Monitoring & Logging / garak.logs \\
\rowcolor{gray!10} 4. Transparency \& Accountability Controls & 4.2 Risk Disclosure & Structured Reporting / garak.report \\

\midrule
\rowcolor{gray!30}
\multicolumn{3}{l}{\textbf{Tool name: Giskard}} \\
\cmidrule{1-3}
\textbf{Broad Category} & \textbf{Taxonomy ID \& Name} & \textbf{Mapping} \\
\cmidrule{1-3}
1. Governance \& Oversight Controls & 1.2 Risk Management & giskard.scan \\
\rowcolor{gray!10} 1. Governance \& Oversight Controls & 1.7 Societal Impact Assessment & giskard.scan logic (Stereotypes \& discrimination) \\
2. Technical \& Security Controls & 2.2 Model Alignment & Groundedness / giskard.scan logic \\
\rowcolor{gray!10} 2. Technical \& Security Controls & 2.3 Model Safety Engineering & giskard.scan logic (Prompt injection/Robustness) \\
2. Technical \& Security Controls & 2.4 Content Safety Controls & giskard.scan logic (Harmful content generation) \\
\rowcolor{gray!10} 3. Operational Process Controls & 3.1 Testing \& Auditing & Systematic Safety Testing and Scenario Orchestration / Scenario class \\
3. Operational Process Controls & 3.2 Data Governance & Sensitive Information Disclosure Detection / giskard.scan \\
\rowcolor{gray!10} 3. Operational Process Controls & 3.5 Post-deployment Monitoring & Agent Interaction Logging and Middleware / CompletionMiddleware \\
3. Operational Process Controls & 3.6 Incident Response \& Recovery & Automated Request Retries and Error Policies / RetryPolicy \\
\rowcolor{gray!10} 4. Transparency \& Accountability Controls & 4.1 System Documentation & Automated Vulnerability Assessment Reporting / scan\_results.to\_html \\
4. Transparency \& Accountability Controls & 4.2 Risk Disclosure & Structured Evaluation Result Disclosure / ScenarioResult \\
\rowcolor{gray!10} 5. Corrective \& Restrictive Actions & 5.1 System \& Feature Restrictions & MinIntervalRateLimiter \\

\midrule
\rowcolor{gray!30}
\multicolumn{3}{l}{\textbf{Tool name: Holistic AI}} \\
\cmidrule{1-3}
\textbf{Broad Category} & \textbf{Taxonomy ID \& Name} & \textbf{Mapping} \\
\cmidrule{1-3}
1. Governance \& Oversight Controls & 1.2 Risk Management & classification\_bias\_metrics function \\
\rowcolor{gray!10} 1. Governance \& Oversight Controls & 1.7 Societal Impact Assessment & Societal Bias and Fairness Quantification \\
2. Technical \& Security Controls & 2.1 Model \& Infrastructure Security & Privacy Risk Measurement \\
\rowcolor{gray!10} 2. Technical \& Security Controls & 2.2 Model Alignment & Fairness Alignment Mitigation \\
2. Technical \& Security Controls & 2.3 Model Safety Engineering & AdversarialDebiasing / ExponentiatedGradientReduction \\
\rowcolor{gray!10} 3. Operational Process Controls & 3.1 Testing \& Auditing & Adversarial Robustness Probing / HopSkipJump attack \\
3. Operational Process Controls & 3.2 Data Governance & Pre-processing Disparate Impact Removal / DisparateImpactRemover \\
\rowcolor{gray!10} 4. Transparency \& Accountability Controls & 4.2 Risk Disclosure & Trustworthiness and Bias Visualization / Visualization tools \\

\midrule
\rowcolor{gray!30}
\multicolumn{3}{l}{\textbf{Tool name: Innodata}} \\
\cmidrule{1-3}
\textbf{Broad Category} & \textbf{Taxonomy ID \& Name} & \textbf{Mapping} \\
\cmidrule{1-3}
1. Governance \& Oversight Controls & 1.2 Risk Management & RedLite Benchmark \\
\rowcolor{gray!10} 1. Governance \& Oversight Controls & 1.7 Societal Impact Assessment & rt-inod-bias dataset \\
2. Technical \& Security Controls & 2.3 Model Safety Engineering & rt-inod-jailbreaking dataset \\
\rowcolor{gray!10} 3. Operational Process Controls & 3.1 Testing \& Auditing & Multi-Domain Safety Auditing / RedLite Benchmark \\
3. Operational Process Controls & 3.3 Access Management & Gated Dataset Access Verification / HF\_TOKEN \\

\midrule
\rowcolor{gray!30}
\multicolumn{3}{l}{\textbf{Tool name: LLM Guard}} \\
\cmidrule{1-3}
\textbf{Broad Category} & \textbf{Taxonomy ID \& Name} & \textbf{Mapping} \\
\cmidrule{1-3}
1. Governance \& Oversight Controls & 1.2 Risk Management & scan\_prompt function / llm\_guard\_api \\
\rowcolor{gray!10} 1. Governance \& Oversight Controls & 1.7 Societal Impact Assessment & Bias output scanner \\
2. Technical \& Security Controls & 2.1 Model \& Infrastructure Security & API Authentication and Rate Limiting \\
\rowcolor{gray!10} 2. Technical \& Security Controls & 2.2 Model Alignment & Relevance output scanner \\
2. Technical \& Security Controls & 2.3 Model Safety Engineering & PromptInjection scanner \\
\rowcolor{gray!10} 2. Technical \& Security Controls & 2.4 Content Safety Controls & Toxicity / BanTopics scanners \\
3. Operational Process Controls & 3.1 Testing \& Auditing & Automated Safety Policy Validation / llm\_guard\_api \\
\rowcolor{gray!10} 3. Operational Process Controls & 3.2 Data Governance & PII and Sensitive Information Anonymization / Anonymize scanner \\
3. Operational Process Controls & 3.3 Access Management & API Authentication and Rate Limiting \\
\rowcolor{gray!10} 3. Operational Process Controls & 3.5 Post-deployment Monitoring & Distributed Tracing and Health Monitoring / OpenTelemetry \\
3. Operational Process Controls & 3.6 Incident Response \& Recovery & Policy-Driven Execution Termination / fail\_fast \\
\rowcolor{gray!10} 4. Transparency \& Accountability Controls & 4.1 System Documentation & Interaction Risk Analysis Logging / LOG\_LEVEL \\
5. Corrective \& Restrictive Actions & 5.1 System \& Feature Restrictions & LLMGuardMaliciousPromptException / block logic \\

\midrule
\rowcolor{gray!30}
\multicolumn{3}{l}{\textbf{Tool name: Langfuse}} \\
\cmidrule{1-3}
\textbf{Broad Category} & \textbf{Taxonomy ID \& Name} & \textbf{Mapping} \\
\cmidrule{1-3}
1. Governance \& Oversight Controls & 1.2 Risk Management & runExperiment / Managed Evaluators, \\
\rowcolor{gray!10} 1. Governance \& Oversight Controls & 1.5 Safety Decision Frameworks & RegressionError / CI/CD Accuracy Gates, \\
1. Governance \& Oversight Controls & 1.7 Societal Impact Assessment & Managed Bias and Toxicity Scoring, \\
\rowcolor{gray!10} 2. Technical \& Security Controls & 2.1 Model \& Infrastructure Security & langfuse.mask() / Sensitive Data Redaction, \\
2. Technical \& Security Controls & 2.2 Model Alignment & Faithfulness and AnswerCorrectness Metrics, \\
\rowcolor{gray!10} 2. Technical \& Security Controls & 2.3 Model Safety Engineering & llm-guard Integration / Security Scoring, \\
2. Technical \& Security Controls & 2.4 Content Safety Controls & Managed Toxicity and Prohibited Content Identification, \\
\rowcolor{gray!10} 3. Operational Process Controls & 3.1 Testing \& Auditing & Systematic Experimentation SDK / Annotation Queues, \\
3. Operational Process Controls & 3.2 Data Governance & Data Retention Policies / Input-Output Masking, \\
\rowcolor{gray!10} 3. Operational Process Controls & 3.3 Access Management & RBAC and SSO Enforcement (GitHub Enterprise, Keycloak, WorkOS) \\
3. Operational Process Controls & 3.5 Post-deployment Monitoring & @observe / Analytics Dashboards,, \\
\rowcolor{gray!10} 3. Operational Process Controls & 3.6 Incident Response \& Recovery & Audit Logs / Usage and Spend Alerts, \\
3. Operational Process Controls & 3.7 Incident Investigation & Trace Graph View / Log Level filtering (DEBUG, ERROR), \\
\rowcolor{gray!10} 4. Transparency \& Accountability Controls & 4.1 System Documentation & Prompt Hub / Version Control Audit Trail, \\
4. Transparency \& Accountability Controls & 4.2 Risk Disclosure & Experiment Compare View / Public Trace Links, \\
\rowcolor{gray!10} 4. Transparency \& Accountability Controls & 4.5 Third-Party System Access & Public API and SDKs,, \\
4. Transparency \& Accountability Controls & 4.6 User Rights \& Recourse & Integrated User Feedback Collection, \\
\rowcolor{gray!10} 4. Transparency \& Accountability Controls & 4.7 Training \& Supportive Measures & Langfuse Academy, \\
5. Corrective \& Restrictive Actions & 5.1 System \& Feature Restrictions & Automated PR Blocking for Quality Regressions, \\
\rowcolor{gray!10} 6. Legal, Regulatory \& Enforcement Actions & 6.2 Regulatory Policy \& Legal Mandates & Comprehensive Audit Trails (ISO 27001, SOC 2, GDPR, HIPAA), \\
7. Financial, Economic \& Market Controls & 7.1 Financial, Economic \& Compensation Remedies & Token Cost Tracking / Spend Alerts \\

\midrule
\rowcolor{gray!30}
\multicolumn{3}{l}{\textbf{Tool name: ModelScan}} \\
\cmidrule{1-3}
\textbf{Broad Category} & \textbf{Taxonomy ID \& Name} & \textbf{Mapping} \\
\cmidrule{1-3}
1. Governance \& Oversight Controls & 1.2 Risk Management & ModelScan.scan method \\
\rowcolor{gray!10} 1. Governance \& Oversight Controls & 1.5 Safety Decision Frameworks & CLI Exit Codes \\
2. Technical \& Security Controls & 2.1 Model \& Infrastructure Security & PickleUnsafeOpScan / SavedModelTensorflowOpScan \\
\rowcolor{gray!10} 2. Technical \& Security Controls & 2.3 Model Safety Engineering & \_check\_for\_unsafe\_tf\_keras\_operator \\
3. Operational Process Controls & 3.1 Testing \& Auditing & Continuous Model Security Auditing / CI/CD integration \\
\rowcolor{gray!10} 3. Operational Process Controls & 3.2 Data Governance & Malicious Data Access Detection / unsafe\_tf\_operators \\
4. Transparency \& Accountability Controls & 4.1 System Documentation & Model Component Inventory and SBOM Generation / --sbom \\
\rowcolor{gray!10} 4. Transparency \& Accountability Controls & 4.2 Risk Disclosure & Standardized Severity Reporting / JSONReport \\
5. Corrective \& Restrictive Actions & 5.1 System \& Feature Restrictions & --strict mode \\

\midrule
\rowcolor{gray!30}
\multicolumn{3}{l}{\textbf{Tool name: NeMo Guardrails}} \\
\cmidrule{1-3}
\textbf{Broad Category} & \textbf{Taxonomy ID \& Name} & \textbf{Mapping} \\
\cmidrule{1-3}
1. Governance \& Oversight Controls & 1.2 Risk Management & LLMRails class / generate method \\
\rowcolor{gray!10} 1. Governance \& Oversight Controls & 1.7 Societal Impact Assessment & Aegis plugin / needs\_caution category \\
2. Technical \& Security Controls & 2.1 Model \& Infrastructure Security & Agentic Action Monitoring \\
\rowcolor{gray!10} 2. Technical \& Security Controls & 2.2 Model Alignment & Conversational Intent Steerage (Dialog rails / Colang flows) \\
2. Technical \& Security Controls & 2.3 Model Safety Engineering & jailbreak detection model / injection detection flow \\
\rowcolor{gray!10} 2. Technical \& Security Controls & 2.4 Content Safety Controls & content safety check input/output flows \\
3. Operational Process Controls & 3.1 Testing \& Auditing & Systematic Rail Performance Evaluation / nemoguardrails evaluate \\
\rowcolor{gray!10} 3. Operational Process Controls & 3.2 Data Governance & Sensitive Data Redaction and Masking / mask sensitive data flows \\
3. Operational Process Controls & 3.3 Access Management & Agentic Tool Call Validation / Execution rails \\
\rowcolor{gray!10} 3. Operational Process Controls & 3.5 Post-deployment Monitoring & Distributed OpenTelemetry Tracing / OpenTelemetry adapter \\
3. Operational Process Controls & 3.6 Incident Response \& Recovery & Policy-Driven Execution Termination / bot refuse to respond \\
\rowcolor{gray!10} 4. Transparency \& Accountability Controls & 4.1 System Documentation & Interaction and Activated Rail Logging / GenerationLog \\
5. Corrective \& Restrictive Actions & 5.1 System \& Feature Restrictions & bot refuse to respond / stop statement \\

\midrule
\rowcolor{gray!30}
\multicolumn{3}{l}{\textbf{Tool name: OpenGuardrails}} \\
\cmidrule{1-3}
\textbf{Broad Category} & \textbf{Taxonomy ID \& Name} & \textbf{Mapping} \\
\cmidrule{1-3}
1. Governance \& Oversight Controls & 1.2 Risk Management & /og\_scan command \\
\rowcolor{gray!10} 1. Governance \& Oversight Controls & 1.7 Societal Impact Assessment & S06 scanner (NSFW Content) \\
2. Technical \& Security Controls & 2.1 Model \& Infrastructure Security & AI Security Gateway \\
\rowcolor{gray!10} 2. Technical \& Security Controls & 2.2 Model Alignment & Intent-Action Mismatch Detection \\
2. Technical \& Security Controls & 2.3 Model Safety Engineering & S01 and S02 scanners (Prompt Injection/System Override) \\
\rowcolor{gray!10} 2. Technical \& Security Controls & 2.4 Content Safety Controls & S10 scanner (Off-Topic Drift) \\
3. Operational Process Controls & 3.1 Testing \& Auditing & Continuous Workspace Security Auditing / /og\_autoscan \\
\rowcolor{gray!10} 3. Operational Process Controls & 3.2 Data Governance & PII and Confidential Data Exposure Protection / S07 scanner \\
3. Operational Process Controls & 3.3 Access Management & Agent Permission and Anomaly Monitoring / observationQueries.findAnomalies \\
\rowcolor{gray!10} 3. Operational Process Controls & 3.5 Post-deployment Monitoring & Real-Time Agentic Hour and Risk Tracking / Agent Activity Monitor \\
3. Operational Process Controls & 3.6 Incident Response \& Recovery & Policy-Driven Action Blocking / Policy engine block \\
\rowcolor{gray!10} 4. Transparency \& Accountability Controls & 4.2 Risk Disclosure & Structured Risk Event Dashboarding / Dashboard API \\
5. Corrective \& Restrictive Actions & 5.1 System \& Feature Restrictions & Policy engine block action \\

\midrule
\rowcolor{gray!30}
\multicolumn{3}{l}{\textbf{Tool name: OpenLIT}} \\
\cmidrule{1-3}
\textbf{Broad Category} & \textbf{Taxonomy ID \& Name} & \textbf{Mapping} \\
\cmidrule{1-3}
1. Governance \& Oversight Controls & 1.2 Risk Management & openlit.evals.All / openlit.guard.All \\
\rowcolor{gray!10} 1. Governance \& Oversight Controls & 1.6 Environmental Impact Management & OPENLIT\_COLLECT\_GPU\_STATS \\
1. Governance \& Oversight Controls & 1.7 Societal Impact Assessment & BiasDetector / ToxicityDetector \\
\rowcolor{gray!10} 2. Technical \& Security Controls & 2.1 Model \& Infrastructure Security & GPU Health Diagnostics \\
2. Technical \& Security Controls & 2.2 Model Alignment & openlit.evals.Hallucination class \\
\rowcolor{gray!10} 2. Technical \& Security Controls & 2.3 Model Safety Engineering & openlit.guard.PromptInjection class \\
2. Technical \& Security Controls & 2.4 Content Safety Controls & SensitiveTopic / TopicRestriction classes \\
\rowcolor{gray!10} 3. Operational Process Controls & 3.1 Testing \& Auditing & Programmatic AI Response Evaluation / openlit.evals module \\
3. Operational Process Controls & 3.2 Data Governance & Centralized API Key and Secret Governance / Vault module \\
\rowcolor{gray!10} 3. Operational Process Controls & 3.5 Post-deployment Monitoring & OpenTelemetry-Native Distributed Tracing / openlit.init \\
4. Transparency \& Accountability Controls & 4.2 Risk Disclosure & AI Performance and Risk Dashboarding / Analytics Dashboard \\
\rowcolor{gray!10} 5. Corrective \& Restrictive Actions & 5.1 System \& Feature Restrictions & Real-Time Guardrails implementation \\
7. Financial, Economic \& Market Controls & 7.1 Financial, Economic \& Compensation Remedies & Automated Token-Level Cost Tracking (pricing\_json parameter) [File: openlit Root README.md] \\

\midrule
\rowcolor{gray!30}
\multicolumn{3}{l}{\textbf{Tool name: PrivacyRaven}} \\
\cmidrule{1-3}
\textbf{Broad Category} & \textbf{Taxonomy ID \& Name} & \textbf{Mapping} \\
\cmidrule{1-3}
1. Governance \& Oversight Controls & 1.2 Risk Management & run\_all\_extraction / ModelExtractionAttack \\
\rowcolor{gray!10} 2. Technical \& Security Controls & 2.3 Model Safety Engineering & ModelExtractionAttack / joint\_train\_inversion\_model \\
3. Operational Process Controls & 3.1 Testing \& Auditing & Comprehensive Privacy Red Teaming / ModelExtractionAttack \\
\rowcolor{gray!10} 3. Operational Process Controls & 3.2 Data Governance & Training Data Membership Verification / Membership inference logic \\
4. Transparency \& Accountability Controls & 4.2 Risk Disclosure & Privacy Vulnerability Visualization / save\_inversion\_results \\

\midrule
\rowcolor{gray!30}
\multicolumn{3}{l}{\textbf{Tool name: Promptfoo}} \\
\cmidrule{1-3}
\textbf{Broad Category} & \textbf{Taxonomy ID \& Name} & \textbf{Mapping} \\
\cmidrule{1-3}
1. Governance \& Oversight Controls & 1.2 Risk Management & promptfoo redteam generate / redteam\_run \\
\rowcolor{gray!10} 1. Governance \& Oversight Controls & 1.7 Societal Impact Assessment & Bias Detection Suite \\
2. Technical \& Security Controls & 2.1 Model \& Infrastructure Security & Model Documentation Scanner \\
\rowcolor{gray!10} 2. Technical \& Security Controls & 2.2 Model Alignment & llm-rubric \\
2. Technical \& Security Controls & 2.3 Model Safety Engineering & promptfoo scan-model / Red Team Strategies (GCG) \\
\rowcolor{gray!10} 2. Technical \& Security Controls & 2.4 Content Safety Controls & guardrails assertion / not-guardrails \\
3. Operational Process Controls & 3.1 Testing \& Auditing & Automated Red Teaming / promptfoo redteam run \\
\rowcolor{gray!10} 3. Operational Process Controls & 3.2 Data Governance & PII Detection / pii plugins \\
3. Operational Process Controls & 3.3 Access Management & RBAC (Enterprise) \\
\rowcolor{gray!10} 3. Operational Process Controls & 3.5 Post-deployment Monitoring & Model Drift Detection / schedule cron \\
3. Operational Process Controls & 3.6 Incident Response \& Recovery & Remediation Reports \\
\rowcolor{gray!10} 4. Transparency \& Accountability Controls & 4.1 System Documentation & Model Documentation Scanner \\
4. Transparency \& Accountability Controls & 4.2 Risk Disclosure & Risk Scoring / promptfoo redteam report \\

\midrule
\rowcolor{gray!30}
\multicolumn{3}{l}{\textbf{Tool name: PyRIT}} \\
\cmidrule{1-3}
\textbf{Broad Category} & \textbf{Taxonomy ID \& Name} & \textbf{Mapping} \\
\cmidrule{1-3}
1. Governance \& Oversight Controls & 1.2 Risk Management & RedTeamingAttack / pyrit\_scan \\
\rowcolor{gray!10} 2. Technical \& Security Controls & 2.2 Model Alignment & SelfAskTrueFalseScorer \\
2. Technical \& Security Controls & 2.3 Model Safety Engineering & CrescendoAttack / SkeletonKeyAttack \\
\rowcolor{gray!10} 2. Technical \& Security Controls & 2.4 Content Safety Controls & AzureContentFilterScorer \\
3. Operational Process Controls & 3.1 Testing \& Auditing & Red Teaming Campaigns / Scenario class \\
\rowcolor{gray!10} 3. Operational Process Controls & 3.2 Data Governance & Data Normalization / PyRIT Database \\
3. Operational Process Controls & 3.6 Incident Response \& Recovery & Automated Feedback Loops / ScenarioResult \\
\rowcolor{gray!10} 4. Transparency \& Accountability Controls & 4.2 Risk Disclosure & Performance Scoring / Scorer metrics \\

\midrule
\rowcolor{gray!30}
\multicolumn{3}{l}{\textbf{Tool name: RAGAS}} \\
\cmidrule{1-3}
\textbf{Broad Category} & \textbf{Taxonomy ID \& Name} & \textbf{Mapping} \\
\cmidrule{1-3}
1. Governance \& Oversight Controls & 1.2 Risk Management & evaluate() function / @experiment \\
\rowcolor{gray!10} 1. Governance \& Oversight Controls & 1.7 Societal Impact Assessment & AspectCritic / harmfulness aspect \\
2. Technical \& Security Controls & 2.2 Model Alignment & Faithfulness / AnswerCorrectness metrics \\
\rowcolor{gray!10} 2. Technical \& Security Controls & 2.3 Model Safety Engineering & AspectCritic / Non-Answer Compliance aspect \\
2. Technical \& Security Controls & 2.4 Content Safety Controls & AspectCritic / harmfulness aspect \\
\rowcolor{gray!10} 3. Operational Process Controls & 3.1 Testing \& Auditing & Batch Performance and Safety Auditing / aevaluate() \\
3. Operational Process Controls & 3.2 Data Governance & Entity-Level Context Coverage Verification / ContextEntityRecall \\
\rowcolor{gray!10} 3. Operational Process Controls & 3.5 Post-deployment Monitoring & Production Trace Analysis and Scoring / OpikTracer \\
4. Transparency \& Accountability Controls & 4.1 System Documentation & Evaluation Metadata and Result Persistence / EvaluationResult.to\_pandas() \\
\rowcolor{gray!10} 4. Transparency \& Accountability Controls & 4.2 Risk Disclosure & Standardized Risk Outcome Sharing / results.upload() \\

\midrule
\rowcolor{gray!30}
\multicolumn{3}{l}{\textbf{Tool name: Rebuff}} \\
\cmidrule{1-3}
\textbf{Broad Category} & \textbf{Taxonomy ID \& Name} & \textbf{Mapping} \\
\cmidrule{1-3}
1. Governance \& Oversight Controls & 1.2 Risk Management & detect\_injection function \\
\rowcolor{gray!10} 2. Technical \& Security Controls & 2.3 Model Safety Engineering & detect\_pi\_heuristics.py module \\
3. Operational Process Controls & 3.1 Testing \& Auditing & Automated Security Unit Testing / test\_sdk.py \\
\rowcolor{gray!10} 3. Operational Process Controls & 3.2 Data Governance & Prompt Instruction Leakage Detection / add\_canary\_word \\
3. Operational Process Controls & 3.5 Post-deployment Monitoring & Real-Time User Input Monitoring / detect\_injection \\
\rowcolor{gray!10} 3. Operational Process Controls & 3.6 Incident Response \& Recovery & Automated Violation Flagging / injection\_detected flag \\
5. Corrective \& Restrictive Actions & 5.1 System \& Feature Restrictions & injection\_detected flag \\

\midrule
\rowcolor{gray!30}
\multicolumn{3}{l}{\textbf{Tool name: WildGuard}} \\
\cmidrule{1-3}
\textbf{Broad Category} & \textbf{Taxonomy ID \& Name} & \textbf{Mapping} \\
\cmidrule{1-3}
1. Governance \& Oversight Controls & 1.2 Risk Management & Automated Multi-Task Safety Moderation (WildGuard.classify method) \\
\rowcolor{gray!10} 2. Technical \& Security Controls & 2.2 Model Alignment & Response Refusal and Compliance Verification (response\_refusal logic) \\
2. Technical \& Security Controls & 2.3 Model Safety Engineering & Adversarial Interaction and Jailbreak Detection (WildGuard moderation tools) \\
\rowcolor{gray!10} 2. Technical \& Security Controls & 2.4 Content Safety Controls & Automated Harmful Response Classification (response\_harmfulness logic) \\
3. Operational Process Controls & 3.1 Testing \& Auditing & Systematic Safety Batch Evaluation (WildGuard.classify with batch support) \\
\rowcolor{gray!10} 3. Operational Process Controls & 3.2 Data Governance & Sensitive Individual and Organizational Information Detection / PII/Privacy classification \\
5. Corrective \& Restrictive Actions & 5.1 System \& Feature Restrictions & Guarded Inference Input Filtering (guarded\_inference.py implementation) \\

\end{xltabular}
\end{landscape}

\clearpage
\bibliographystyle{IEEEtran}
\bibliography{references}

\end{document}